\documentclass[aps,pra,twocolumn,preprintnumbers,11pt,final,tightenlines,superscriptaddress,
noshowpacs,nofloats,showkeys,showkeywords,authoryear]{revtex4-2}

\usepackage{graphicx}
\usepackage{amsmath}
\usepackage[T2A]{fontenc}
\usepackage[utf8]{inputenc}
\usepackage[english]{babel}

\usepackage{graphicx}
\usepackage{dcolumn}
\usepackage{bm}

\usepackage{amsmath}
\usepackage{amssymb}
\usepackage[T2A]{fontenc}
\usepackage[utf8]{inputenc}
\usepackage[english]{babel}
\usepackage{multirow}
\usepackage{xcolor}
\usepackage{todonotes}
\usepackage{float}
\usepackage{booktabs}
\usepackage{changes}
\usepackage{longtable}
\allowdisplaybreaks
\setcitestyle{authoryear,round}
\makeatletter

\renewcommand\thetable{\arabic{table}}

\makeatother

\def\hm{HM\,Sge}

\def\arcsec{\hbox{$^{\prime\prime}$}}
\DeclareRobustCommand{\ion}[2]{\textup{#1\,\textsc{\lowercase{#2}}}}

\graphicspath{{figs/}}

\begin{document}
\selectlanguage{english}

\title{Photometric and Spectral Evolution \\of the Symbiotic Nova HM Sagittae since 2003}

\author{N.P.~Ikonnikova}
\email{ikonnikova@gmail.com}
\affiliation{Sternberg Astronomical Institute, M.V. Lomonosov Moscow State University, Moscow, 119234 Russia}

\author{V.I.~Shenavrin}
\affiliation{Sternberg Astronomical Institute, M.V. Lomonosov Moscow State University, Moscow, 119234 Russia}

\author{G.V.~Komissarova}
\affiliation{Sternberg Astronomical Institute, M.V. Lomonosov Moscow State University, Moscow, 119234 Russia}

\author{M.A.~Burlak}
\affiliation{Sternberg Astronomical Institute, M.V. Lomonosov Moscow State University, Moscow, 119234 Russia}

\author{A.V.~Dodin}
\affiliation{Sternberg Astronomical Institute, M.V. Lomonosov Moscow State University, Moscow, 119234 Russia}

%\date{\today}

\begin{abstract}

We present photometric and spectral monitoring of the symbiotic system \hm, which consists of a Mira variable with a dust shell and a hot white dwarf ionizing the surrounding gaseous nebula. The system underwent a nova-like outburst in 1975 and experienced a high-activity episode during 2018--2021. $UBV$ photometry from 2003 to 2025 shows monotonic fading at $\sim0^{m}.05$~yr$^{-1}$ until 2018, followed by a $0^{m}.3$ brightening peaking in 2021 and a decline by 2025 to the faintest level in five decades of monitoring. Near-infrared observations ($JHKLM$, 2009--2025), combined with archival data, reveal Mira pulsations with a period of 532~d and long-term variability driven by changes in the optical depth of the dust shell. Spectral monitoring (2016--2025) reveals a substantial evolution in the emission spectrum of the gaseous envelope. The 2018--2021 high-activity episode was accompanied by enhanced fluxes in recombination lines (\ion{H}{I}, \ion{He}{I}, \ion{He}{II}) and forbidden transitions ([\ion{O}{I}], [\ion{Ar}{V}], [\ion{Fe}{VI}], [\ion{Ca}{VII}]), together with a 17-fold brightening of the Raman-scattered \ion{O}{VI}~$\lambda$6725 line. We report the first detection of the [\ion{Fe}{X}]~$\lambda$6374 line in \hm. Monitoring this line from 2007 to 2025 shows its equivalent width growing through 2017 -- indicating gradual coronal heating -- then declining by 2021, likely reflecting altered accretion conditions and/or hot-component properties during the high-activity episode. We propose that both the 1975 outburst and the 2018--2021 high-activity episode may be linked to periastron passage of the binary components; if so, the $\sim46$-year interval would constrain the system's orbital period.

\end{abstract}
\maketitle

\section{Introduction}

Symbiotic novae constitute a unique class of interacting binary systems in which a thermonuclear explosion on the surface of an accreting white dwarf (WD) leads to a long-lasting (from several decades to centuries) and substantial increase in luminosity. Investigation of the long-term post-outburst evolution provides essential insights into mass-loss mechanisms, the stability of nuclear burning, and the physical conditions within the complex multi-component environment surrounding the binary system.

Only about ten confirmed objects in the Galaxy are known to have exhibited nova-like outbursts over the past century and a half. Among the most prominent and well-studied are V1016\,Cyg, V1329\,Cyg, AG\,Peg, and RR\,Tel. \hm{} occupies a special place among such objects -- a D-type (Dusty) symbiotic system according to the classification of \cite{allen1984}, containing a Mira-type variable surrounded by a dust shell. Fifty years ago, in 1975, this system underwent a powerful nova-like outburst \citep{dokuchaeva1976}, which brought it out of quiescence and established it as one of the primary targets for studying the late evolutionary stages of symbiotic novae.

Subsequent decades of observations have made it possible to trace in detail the gradual fading of the outburst and the evolution of the system's physical parameters. However, in the late 2010s, \hm\ once again exhibited signs of increased activity. Observations from 2018--2021 recorded a brightening and spectral changes indicative of a possible resumption of thermonuclear activity, a change in the accretion regime, or a large-scale restructuring of the circumstellar environment. This event, which occurred half a century after the main outburst, raises fundamental questions about the nature of the late evolutionary stages of symbiotic novae, the mechanisms of outburst recurrence, and the long-term stability of nuclear burning on WDs.

In this work, we present a comprehensive analysis of observational data for \hm, covering both the entire 50-year period since the 1975 outburst and the recent activity of 2018--2021. Using new photometric and spectral observations, supplemented by archival data, we investigate the evolution of the ionized nebula and dust shell surrounding the system.

Some key parameters of the system are summarized in Table~\ref{tab:hmsge_params}.

\begin{table*}[htbp]
\centering
\caption{Key parameters of the symbiotic nova \hm. 
}
\label{tab:hmsge_params}
\begin{tabular*}{\textwidth}{@{\extracolsep{\fill}} l l l}
\toprule
Parameter & Value & Reference \\
\midrule
\multicolumn{3}{l}{\textbf{Coordinates and distance:}} \\
Right Ascension (RA, J2000) & $19^{\mathrm{h}}41^{\mathrm{m}}57.1^{\mathrm{s}}$ & \cite{gaia2021}\\
Declination (Dec, J2000) & $+16^{\circ} 44' 39.8''$ & \cite{gaia2021}\\
Distance & $1035^{+118}_{-102}$ pc & \cite{bailerjones2021} \\
\midrule
\multicolumn{3}{l}{\textbf{Mira-type star (cool companion):}} \\
Spectral class & M7 & \cite{muerset1999}\\
Luminosity & $\sim1000~L_{\odot}$ & \cite{taranova2000}\\
& $1500-2000~L_{\odot}$ & \cite{goldman2024}\\
Radius & $\sim540~R_{\odot}$ & \cite{taranova2000}\\
Temperature & 2400--3000~K & \cite{taranova2000}\\
Mass-loss rate (gas) & $\sim4 \times 10^{-6}~M_{\odot}$\,yr$^{-1}$ & \cite{goldman2024}\\
Pulsation period & $535\pm5$ days & \cite{taranova2000}\\
Pulsation amplitude ($\Delta J$) & $\sim1.^{m}0$ &\cite{taranova2000}\\
\midrule
\multicolumn{3}{l}{\textbf{White dwarf (hot component):}} \\
Temperature (2021) & $>250\,000$~K & \cite{goldman2024} \\
\midrule
\multicolumn{3}{l}{\textbf{Dust shell:}} \\
Radius & $\sim1500~R_{\odot}$ & \cite{taranova2000}\\
Mass & $4-8\times10^{-5}~M_{\odot}$ & \cite{taranova2000}\\
\bottomrule
\end{tabular*}
\end{table*}

\section{$UBV$ observations}

The results of $UBV$ observations of \hm{} obtained prior to 2002 have been published previously \citep{arkhipova1994, arkhipova2004}. In recent years, new observations have been carried out using the same 60-cm reflector at the Crimean Astronomical Station (CAS) of SAI MSU, equipped with the photometer 
designed by V.\,M.~Lyuty \citep{lyuty1971} and operating in the same photometric system. Observations were performed using a 13\arcsec{} diaphragm to exclude the contribution of a faint companion located 8\arcsec{} north of \hm. As in previous studies, BD+16$^{\circ}$3965 (spectral class F5) was used as the comparison star. The magnitudes of the comparison star were adopted as $V = 10^{m}.70$, $B = 11^{m}.33$, and $U = 11^{m}.48$ \citep{arkhipova1994}. Our $UBV$ observations of \hm{} from 2003 to 2025 are presented in Table~\ref{tab:UBV}. We estimate the photometric precision to be $0.^{m}01-0.^{m}02$ in the $V$ and $B$ bands, and no worse than $0.^{m}03$ in the $U$ band.

\section{$JHKLM$ observations}

$JHKLM$ photometry of \hm{} has been carried out on the CAS SAI MSU 1.25-m telescope since 1978. Infrared (IR) flux measurements were carried out using an InSb photometer developed and built in the early 1980s \citep{nadjip1986}. A detailed description of the observing technique and data reduction is given in \cite{shenavrin2011}. The star BS~7488 from the catalog of \cite{johnson1966} is used as the photometric standard. The angular diameter of the exit diaphragm is approximately 12\arcsec, and the spatial beam separation during modulation is $\sim 30\arcsec$ in the east--west direction. Measurement uncertainties do not exceed $0.^{m}01$ in the $J$, $H$, $K$, and $L$ bands, and $0.^{m}02$ in the $M$ band. The results of $JHKLM$ photometry from 1978 to 2008 have been published in \cite{taranova1983, yudin1994, yudin1994a, taranova2000, shenavrin2011}. Our near-IR observations obtained during 2009--2025 are presented in Table~\ref{tab:JHKLM}.

\section{Spectroscopic observations}

Spectroscopic observations of \hm{} during 2016--2025 have been carried out on the CAS SAI MSU 1.25-m telescope using the low-resolution spectrograph (A-sp) designed by V.\,F.~Esipov, with a 600~lines\,mm$^{-1}$ diffraction grating, a 4\arcsec-wide long slit, $\mathrm{FWHM} \sim 7$~\AA, and a spectral range of 4000--10000~\AA. The following detectors were used: the SBIG ST-402 CCD (765$\times$510 pixels) in 2016--2017, the FLI PL-4022 (2048$\times$2048 pixels) in 2021, and the CMOS camera ASI2600MM/MC Pro (6248$\times$4176 pixels) in 2024--2025. Spectroscopic observations of \hm{} on the same telescope during 2003--2016 were also obtained; however, due to issues with absolute flux calibration of spectra, we present only equivalent widths for that period.

In 2021 and 2025, additional spectroscopic observations of \hm{} were obtained on the CMO SAI MSU 2.5-m telescope using the low-resolution Transient Double-beam Spectrograph (TDS) equipped with holographic gratings \citep{potanin2020}. Andor Newton 940P cameras with E2V CCD42-10 detectors (512$\times$2048 pixels) were used. Observations were performed with a long slit of width 1.\arcsec0{} or 1.\arcsec5. The covered spectral range is 3500--7500~\AA, with spectral resolving power $R \approx 1300$ in the blue channel (3500--5720~\AA) and $R \approx 2500$ in the red channel (5720--7500~\AA). Data reduction was performed using custom Python scripts; the procedure is described in detail in \cite{potanin2020}.

The log of spectroscopic observations is presented in Table~\ref{tabl_sp}.

\begin{table}
\centering
    \caption{Log of spectroscopic observations.}  \label{tabl_sp}
    \begin{tabular}{cccc}
     \hline
      Date   &  JD-2400000 & Instrument  & Standard \\
      \hline
    2016-07-30 &57600.46 & A-sp & HD 353437\\
    2017-07-01 &57936.45 & A-sp & HD 353437\\
    2021-06-10 &59376.45 & TDS & BD+33$^{\circ}$ 2642\\
    2021-07-15 &59411.33 & A-sp & HD 353437\\
    2021-09-09 &59467.33 & A-sp & HD 353437\\
    2024-07-07 &60499.41 & A-sp & HD 353437\\
    2025-05-04 &60800.47 & TDS & HIP 87939\\
    2025-08-14 &60902.38 & A-sp & HD 353437\\ 
    2025-11-18 &60998.18 & TDS & HIP 106758\\
      \hline
    \end{tabular}
\end{table}

\section{Photometric evolution}

\subsection{Optical range}

\begin{figure*}
   \centering
   \includegraphics[width=\hsize]{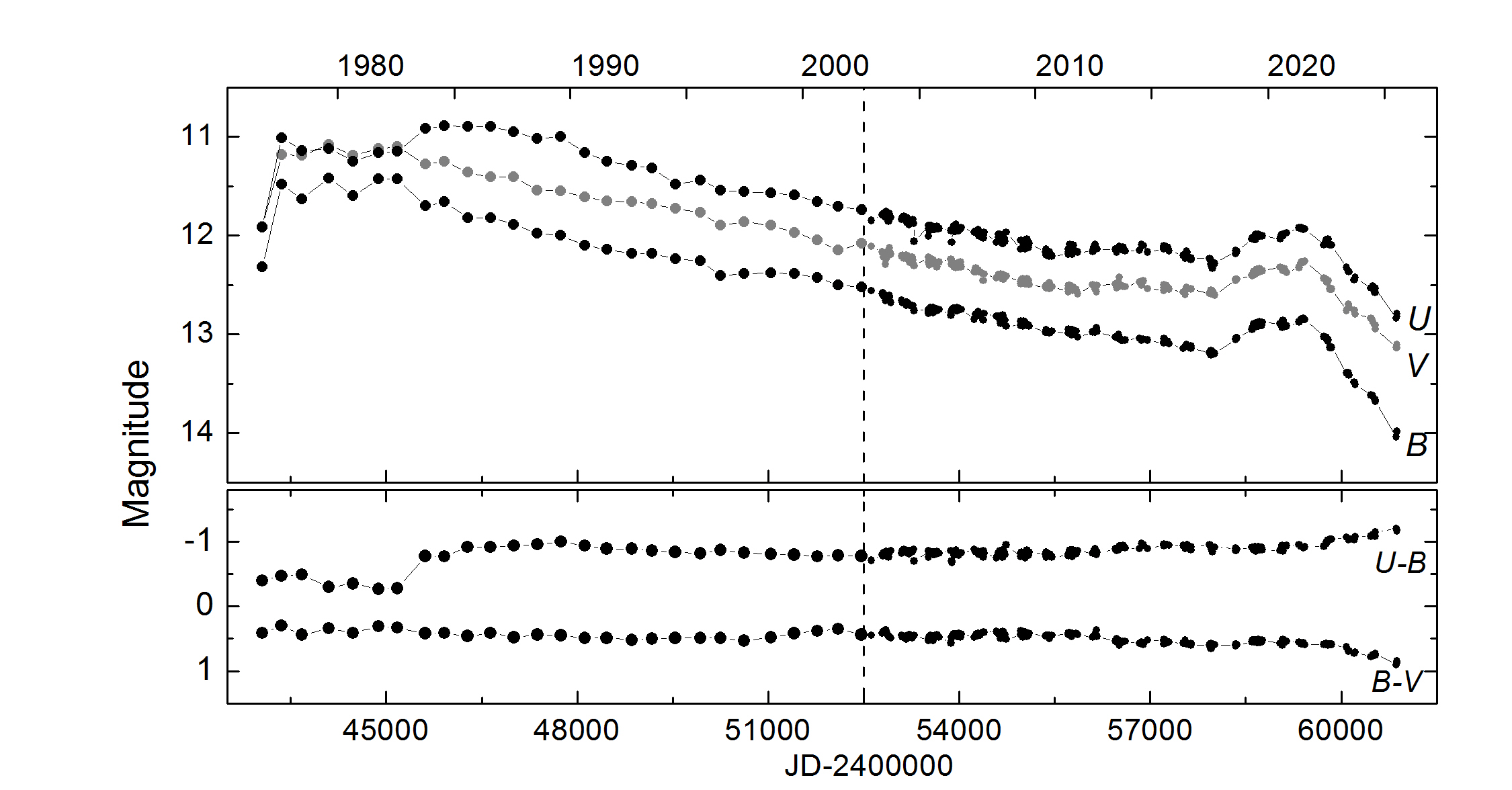}
\caption{Light and color curves of \hm{} from 1976 to 2025. New data are shown to the right of the vertical dashed line. 
}

\label{fig1}
   \end{figure*}

The light curves covering the entire period of $UBV$ observations are shown in Fig.\,\ref{fig1}, which displays the annual mean magnitudes for 1976--2002 from \cite{arkhipova2004} and new data obtained during 2003--2025. As noted previously \citep{arkhipova1994}, the mean brightness in all three bands remained constant from 1977 to 1982, with fluctuations of amplitude $\sim 0.^{m}1$, followed by a brightening in the $U$ band. \hm{} remained in this ``high'' state for approximately 7 years, and starting in 1989, a systematic fading occurred at a rate of $\sim 0.^{m}05$\,yr$^{-1}$ until 2017. In the $B$ and $V$ bands after 1982, the brightness declined monotonically at approximately the same rate as in the $U$ band. Intra-seasonal brightness variations did not exceed $0.^{m}1$--$0.^{m}2$.

% --  --  --  --  --  --  --  -- -Fig. 1 --  --  --  --  --  --  --  -- 

% --  --  --  --  --  --  --  --  --  --  --  --  --  --  --  -- - 

In 2018, the brightness of \hm{} began to increase in all observed bands, reaching a maximum in 2021 with an amplitude of $\sim 0.^{m}3$ relative to the 2017 level. In subsequent years, a monotonic fading was observed, and by 2025 the brightness reached the minimum values recorded in the last 50 years. If the fading trend observed before 2018 were extrapolated, the expected brightness level in 2025 would have been higher than observed. This indicates that the 2018--2021 event significantly changed the state of the system, 
preventing the hot star and its ionized envelope from returning to the pre-brightening state.

\begin{figure}
   \centering
   \includegraphics[width=8cm]{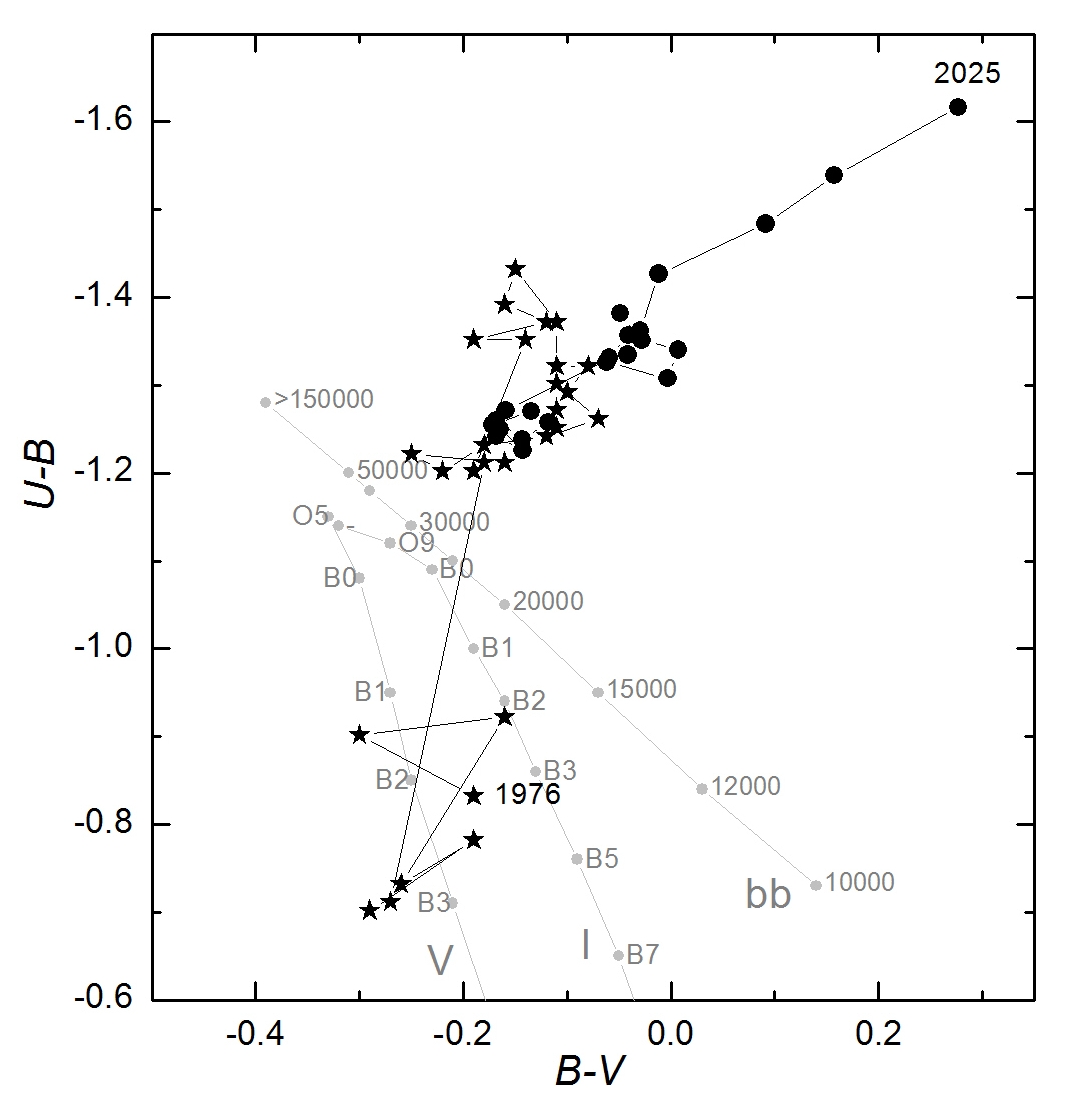}
\caption{Motion of \hm{} in the two-color diagram. Annual mean values are shown, corrected for interstellar extinction with $E(B-V) = 0.6$. Years are indicated by numbers. Data from 1976--2002 are shown as stars, and from 2003 to 2025 as circles. Gray lines show the loci of blackbody radiation (bb), main-sequence stars (V), and supergiants (I). }

\label{fig2}
   \end{figure}

Previously, the behavior of \hm{} in the $(B-V)$, $(U-B)$ two-color diagram was discussed in \cite{arkhipova1994, arkhipova2004}. Here we present an updated two-color diagram (Fig.~\ref{fig2}), extended with observations from the last two decades. In recent years, the previously identified trend has persisted: the star continues to shift toward bluer $(U-B)$ values while simultaneously reddening in $(B-V)$. As in previous studies, we attribute this motion in the diagram to changes in the parameters of the gaseous envelope, which remains the dominant source of emission in the $UBV$ photometric bands. The contributions of both the hot star and the cool component (the Mira variable) are substantially smaller in this spectral region. As will be shown below (see the section on spectral evolution), the emission-line spectrum of the system has undergone significant changes, further confirming the dynamic nature of the observed processes.

% --  --  --  --  --  --  --  -- -Fig. 2 --  --  --  --  --  --  --  -- 

% --  --  --  --  --  --  --  --  --  --  --  --  --  --  --  -- - 

\subsection{Near-IR range}

\begin{figure*}
   \centering
   \includegraphics[width=\hsize]{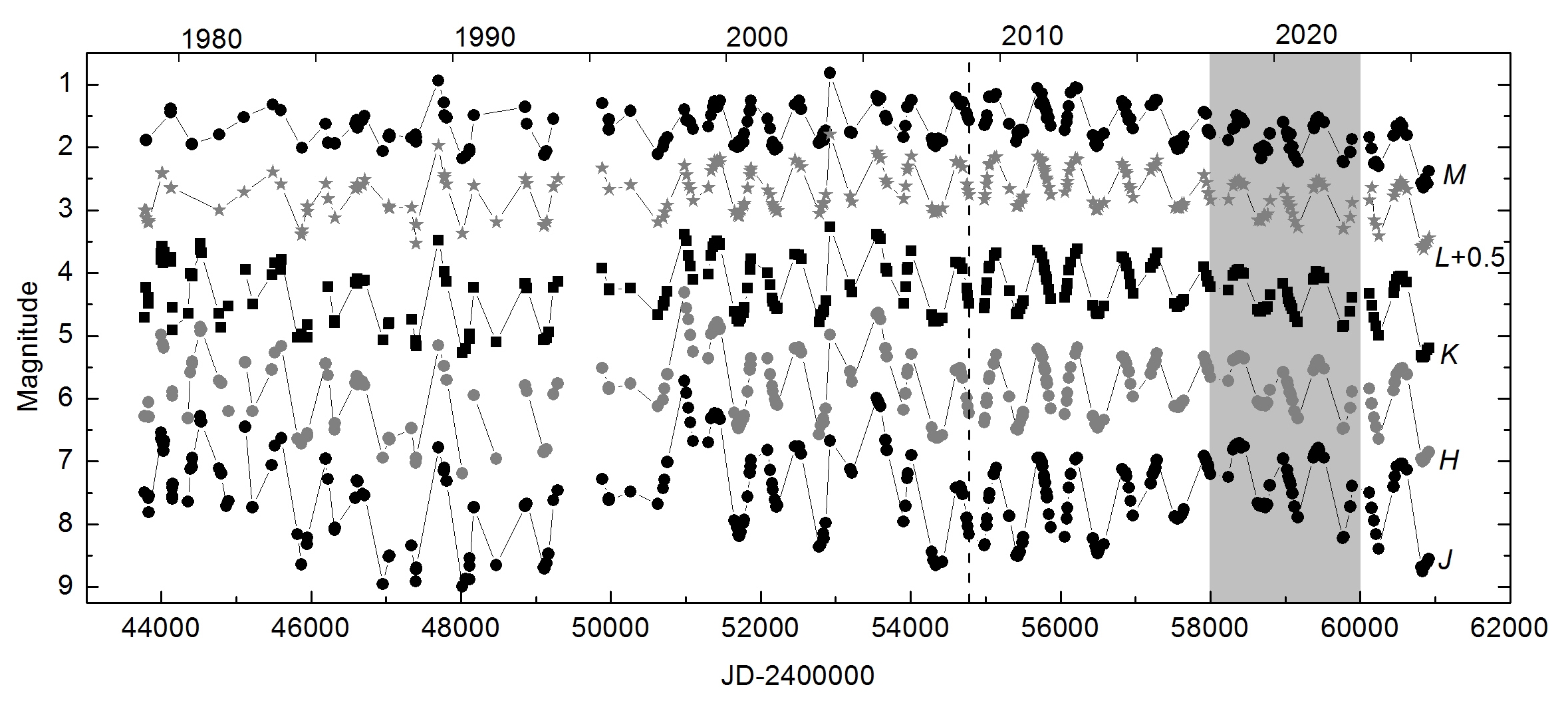}
\caption{Light curves in the $JHKLM$ bands from 1978 to 2025. New data are shown to the right of the vertical dashed line. The period of optical brightening is highlighted in gray. }

\label{fig3}
   \end{figure*}

Figure~\ref{fig3} presents the light curves of \hm{} in the $JHKLM$ bands from 1978 to 2025. New data from 2009--2025 are shown to the right of the vertical dashed line. Earlier observations are taken from \cite{taranova1983, yudin1994, yudin1994a, taranova2000, shenavrin2011}. The time interval during which optical brightening was recorded is highlighted in the figure. It can be seen that only in the $J$ band did the brightness increase slightly during this period.

% --  --  --  --  --  --  --  -- -Fig. 3 --  --  --  --  --  --  --  -- 

% --  --  --  --  --  --  --  --  --  --  --  --  --  --  --  -- - 

\begin{figure*}
   \centering
   \includegraphics[width=\hsize]{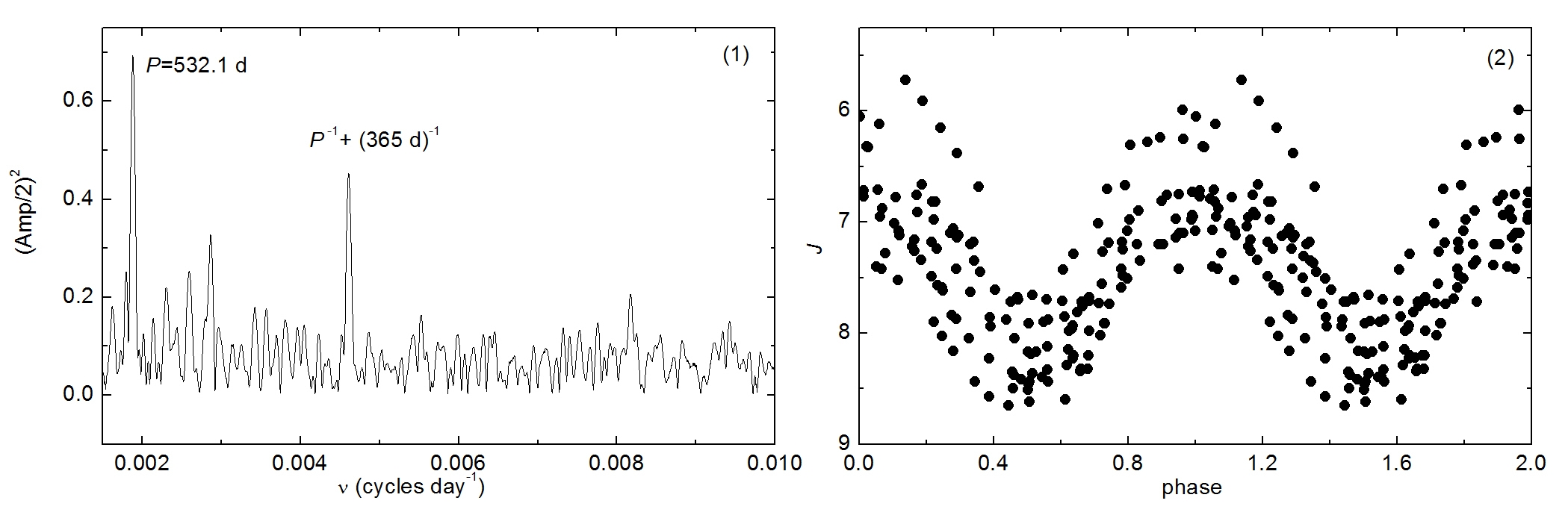}
\caption{Periodogram of the $J$-band time series (1) and the corresponding phase-folded light curve with $P = 532.1$\,d (2), based on the complete 1978--2025 dataset.}

\label{fig4}
   \end{figure*}

The star exhibits stable periodic variations superimposed on a long-term non-monotonic trend. The peak-to-peak amplitudes of periodic variations in different photometric bands are: $\Delta J = 1.^{m}4$, $\Delta H = 1.^{m}2$, $\Delta K = 0.^{m}95$, $\Delta L = 0.^{m}8$, and $\Delta M = 0.^{m}7$. Figure~\ref{fig4} shows the periodogram constructed from the $J$-band data spanning 1978--2025. The derived period $P = 532 \pm 5$~days agrees well (within uncertainties) with the value obtained previously from a shorter time series \citep{taranova2000}.

% --  --  --  --  --  --  --  -- -Fig. 4 --  --  --  --  --  --  --  -- 

% --  --  --  --  --  --  --  --  --  --  --  --  --  --  --  -- - 

\begin{figure}
   \centering
   \includegraphics[width=\hsize]{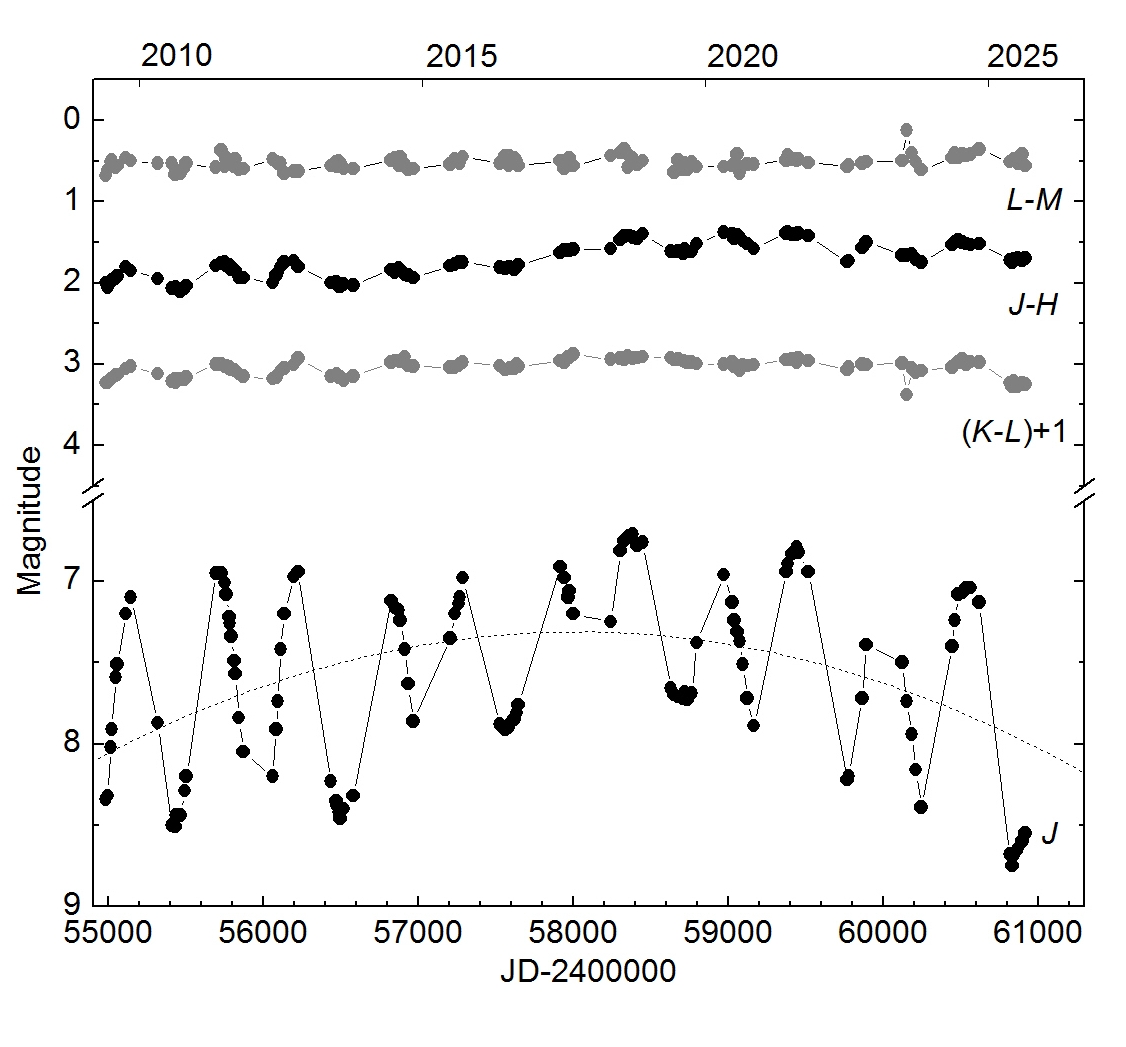}
\caption{Light and color curves from 2008 to 2025. The dashed line shows a second-order polynomial fit. }

\label{fig5}
   \end{figure}

In addition to periodic brightness variations, slow variations with characteristic timescales of several thousand days have been detected in the near-IR range. The most pronounced trend is observed in the $J$ band, which may be related to changes in the optical depth of the star's dust shell. According to \cite{taranova2000}, following the 1975 outburst of the hot component, the dust shell reached maximum density approximately 13 years later (around JD~2447500), after which its gradual dispersal began. By JD~2451000, the $J$-band brightness reached a maximum and then started to decline until JD~2454500; after this date, the mean brightness began to increase again. Another decline in mean brightness is observed following the 2018--2021 event. Early observations in 2025 show that the star's brightness has decreased in all NIR bands to the level of the early 1990s, when the dust shell was at its maximum density. One might hypothesize that a dust-formation episode occurred in the system, associated with a weak recurrent outburst; however, the expected reddening of color indices was not observed. Figure~\ref{fig5} shows the light curve in the $J$ band and the $J-H$, $K-L$, and $L-M$ color curves for the recent period from 2008 to 2025. Long-term brightness variations are fitted with a quadratic polynomial. The $J-H$ color index exhibits bluer values starting in 2018 (the onset of optical brightening), while the $J$-band brightness fades. The color indices $K-L$ and $L-M$ have not undergone significant changes.

% --  --  --  --  --  --  --  -- -Fig. 5 --  --  --  --  --  --  --  -- 

% --  --  --  --  --  --  --  --  --  --  --  --  --  --  --  -- - 

\begin{figure}
   \centering
   \includegraphics[width=\hsize]{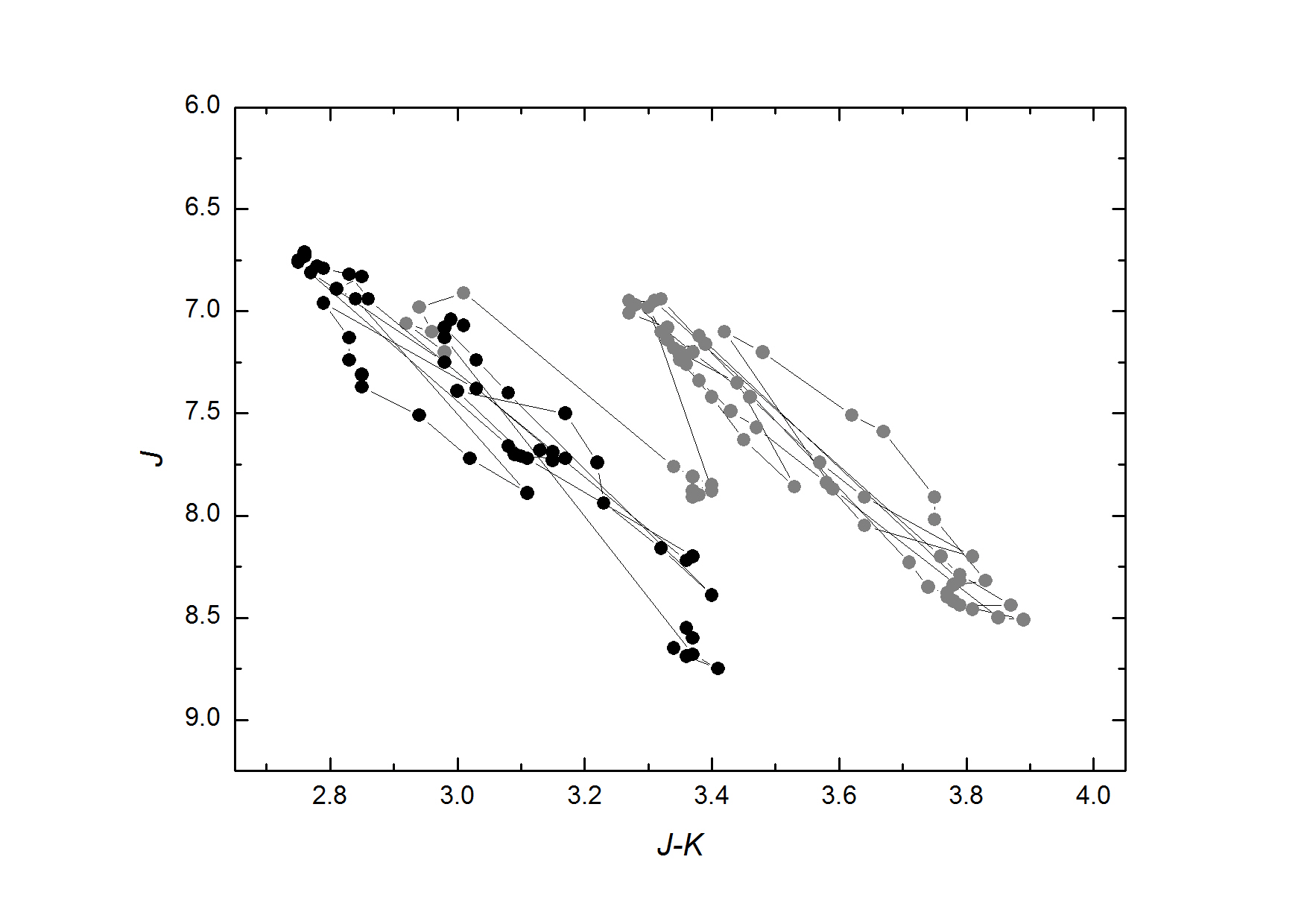}
\caption{Color--magnitude diagram ($(J-K)$ vs.\ $J$) for 2009--2025. Gray points show data from 2008--2017; black points show data from 2018--2025.}

\label{fig6}
   \end{figure}

The $(J-K)$ vs.\ $J$ color--magnitude diagram (Fig.~\ref{fig6}) is of particular interest. Recent observations clearly separate into two temporal groups: 2008--2018 and 2019--2025. Within each group, the star reddens as it fades during pulsations; however, the mean color index has become bluer in 2019--2025, indicating that the system parameters changed during and after the high-activity episode.

% --  --  --  --  --  --  --  -- -Fig. 6 --  --  --  --  --  --  --  -- 

% --  --  --  --  --  --  --  --  --  --  --  --  --  --  --  -- - 

\section{Spectral evolution}

Over half a century of observations, the optical spectrum of the symbiotic nova \hm{} has undergone significant changes, reflecting the evolution of its hot component. The variable flux of ionizing radiation has affected the state of the surrounding gaseous envelope, which is the dominant source of emission in the optical range. The spectrum was studied most extensively during the first two decades after the 1975 outburst. Thus, monitoring results from 1976 and 1977 are presented in \cite{ciatti1977, ciatti1978}, and the evolution of the emission-line spectrum from 1977 to 1982 was traced by \cite{stauffer1984}. Data from 1985--1986 were summarized by \cite{pacheco1989}, and observations from 1987 were analyzed by \cite{schmid1990}. In addition, analysis of spectra from individual epochs is contained in \cite{davidson1978, brown1978, shanin1978, puetter1978, arkhipova1979, ipatov1981, blair1981}. A brief review of spectroscopic observations during the first decade after the outburst is presented in \cite{nussbaumer1990}. Later observations are reported in a series of papers \citep{arkhipova1991, arkhipova1994, noskova1996, arkhipova2004}. Here we analyze new spectroscopic observations of \hm\ obtained during 2016--2025 and compare them with previously published data.

The observed optical spectrum of \hm\ represents a complex superposition of emission components arising from regions with vastly different physical conditions. The spectrum is rich in emission lines. The most intense among them are hydrogen Balmer series lines, forbidden [\ion{O}{III}] lines, \ion{He}{II} lines, and highly ionized [\ion{Fe}{VII}] lines. Weaker lines of \ion{He}{I}, [\ion{O}{I}], [\ion{O}{II}], [\ion{N}{II}], [\ion{S}{II}], [\ion{S}{III}], [\ion{Ar}{III}], [\ion{Ar}{V}], [\ion{Ca}{V}], and [\ion{Fe}{VI}] are also observed. Such a spectrum is characteristic of a gaseous nebula excited by radiation from a high-temperature source. In the red spectral region ($\lambda > 6000$~\AA), the continuum of the red component (the Mira variable) is present, with the intensity varying significantly depending on the pulsation cycle phase \citep{arkhipova2004}.

In the first years after the 1975 outburst, the most significant changes in the emission-line spectrum of \hm\ were observed during a 10-month interval beginning in August 1979, as described in detail by \cite{stauffer1984}. Before August 1979, the effective temperature of the hot component was below 70\,000~K, and the spectrum showed signatures of a high-velocity wind -- Wolf--Rayet bands and broad Balmer line wings. By May 1980, the temperature of the hot star had increased to $\sim$160\,000~K, the wind had virtually disappeared, and [\ion{Fe}{VII}] lines appeared in the spectrum.

In the following decades, the absolute fluxes of nearly all emission lines gradually declined \citep{arkhipova1991, muerset1994, arkhipova2004}. Such fading is characteristic of symbiotic novae and is a natural consequence of the cooling, expansion, and recombination of the ejected envelope. This process spans decades, consistent with the characteristic timescales of cooling and expansion in the dense environment of symbiotic systems. The temperature of the hot star continued to rise, reaching $\sim$200\,000~K by 1992 \citep{muerset1994}. \cite{goldman2024} reported that the temperature of the hot component, based on ultraviolet spectroscopic observations from 2021, exceeds 250\,000~K, derived from the flux ratio of the [\ion{Ne}{V}]~$\lambda$1575 and [\ion{Ne}{IV}]~$\lambda$1601 lines.

The 2018--2021 high-activity episode dramatically altered the emission-line spectrum, disrupting the long-term spectral evolution.

\subsection{Variation of absolute emission-line fluxes}

Observations from 2016 revealed a decline in fluxes of nearly all measured emission lines compared to the data from \cite{arkhipova2004} for 1994--2002. The only exception was the Raman-scattered \ion{O}{VI}~$\lambda$6825 line, whose flux remained almost unchanged relative to the 2002 measurements. The degree of line fading varied depending on the ion. Thus, the H$\beta$ and \ion{He}{II} lines faded by a factor of $\sim$2.4. The nebular [\ion{O}{III}]~$\lambda\lambda$4959, 5007 lines weakened by a factor of $\sim$3.5, while the auroral [\ion{O}{III}]~$\lambda$4363 and [\ion{N}{II}]~$\lambda$5755 lines faded by more than a factor of 4. The [\ion{S}{II}]~$\lambda$6716 and $\lambda$6731 lines proved most stable, showing minimal fading. At the same time, the flux of the [\ion{S}{III}]~$\lambda$6312 line decreased by nearly a factor of 4. The \ion{He}{I}~$\lambda\lambda$5876, 6678, and 7065 lines became weaker by a factor of 2--2.5.

\begin{figure*}
   \centering
   \includegraphics[width=\hsize]{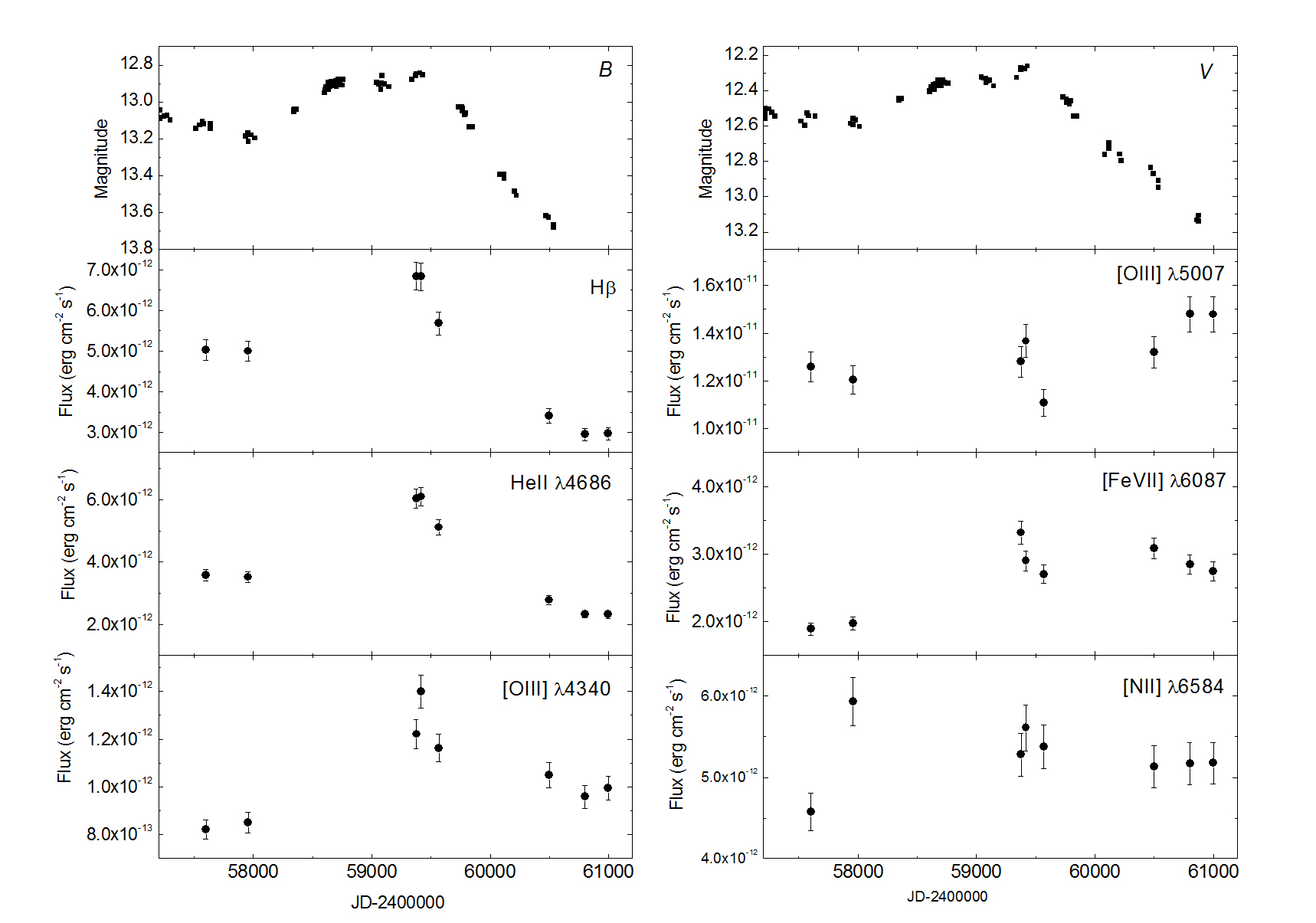}
\caption{Evolution of absolute fluxes for selected emission lines from 2016 to 2025, together with $B$- and $V$-band light curves.}

\label{fluxes}
   \end{figure*}

Our spectroscopic observations have allowed us to trace the evolution of emission-line fluxes during 2016--2025, a period that covered the phases before, during, and after the 2018--2021 high-activity episode. Tables~\ref{tab:ZTE} and \ref{tab:TDS} present absolute line fluxes measured from spectra obtained with the CAS SAI MSU 1.25-m and CMO SAI MSU 2.5-m telescopes. Figure~\ref{fluxes} shows the evolution of absolute fluxes for selected emission lines from 2016 to 2025 based on our new observations, together with $B$- and $V$-band light curves for the same period.

% --  --  --  --  --  --  --  -- -Fig. 7 --  --  --  --  --  --  --  -- 

% --  --  --  --  --  --  --  --  --  --  --  --  --  --  --  -- - 

The H$\beta$ flux increased by a factor of 1.4 from 2016 to 2021, then began to decline and by 2025 reached its minimum value throughout the history of spectroscopic observations of \hm\ -- $\sim3\times10^{-12}$~erg~cm$^{-2}$~s$^{-1}$. The \ion{He}{II}~$\lambda$4686 line showed similar behavior: its absolute flux increased from 2016 to 2021 and then decreased by 2025. The maximum luminosity of the hot component during the outburst led to the highest degree of ionization and, consequently, to the maximum intensity of recombination lines and high-excitation lines. In addition, we note the presence of broad H$\alpha$ wings in the 2021 spectrum and their absence in 2025. The flux of the auroral [\ion{O}{III}]~$\lambda$4363 line increased by a factor of 1.7 from 2016 to 2021, then began to decline; in 2025, it remained slightly higher than in 2016. A different pattern was observed for the nebular lines [\ion{O}{III}]~$\lambda$5007 and [\ion{N}{II}]~$\lambda$6584. The [\ion{O}{III}]~$\lambda$5007 line showed little response to the outburst, but its flux increased slightly afterward. Presumably, the ionization wave from the outburst reached more distant and diffuse regions of the envelope, where this line forms efficiently. The [\ion{N}{II}]~$\lambda$6584 line showed relative stability: its flux did not experience significant enhancement during the photometric maximum of the star. This is consistent with the view that its formation region during this period was only weakly affected by processes related to the outburst.

Lines showing behavior similar to H$\beta$ include \ion{He}{I}~$\lambda\lambda$4471, 5876, 6678, 7065, \ion{He}{II}~$\lambda$5412, and the forbidden lines [\ion{O}{I}]~$\lambda\lambda$6300, 6364, [\ion{Ar}{V}]~$\lambda\lambda$6435, 7006, [\ion{Fe}{VI}]~$\lambda$5158, and [\ion{Ca}{VII}]~$\lambda$5716. The high-excitation lines [\ion{Fe}{VII}]~$\lambda\lambda$5721, 6087 increased in strength by 2021 and maintained enhanced intensity through 2025. For the first time in the history of observations of \hm, the [\ion{Fe}{X}] line was detected, whose emergence and evolution is described in Section~6.1.2. The [\ion{O}{II}]~$\lambda\lambda$7320, 7331 doublet showed little response to the episode: the combined flux of these lines gradually declined from 2016 to 2025. Significant enhancement during the high-activity episode was exhibited by the permitted \ion{Fe}{II}~$\lambda\lambda$4657, 5317, 5535, 6417, 6455 lines.

\begin{figure*}
   \centering
   \includegraphics[width=\hsize]{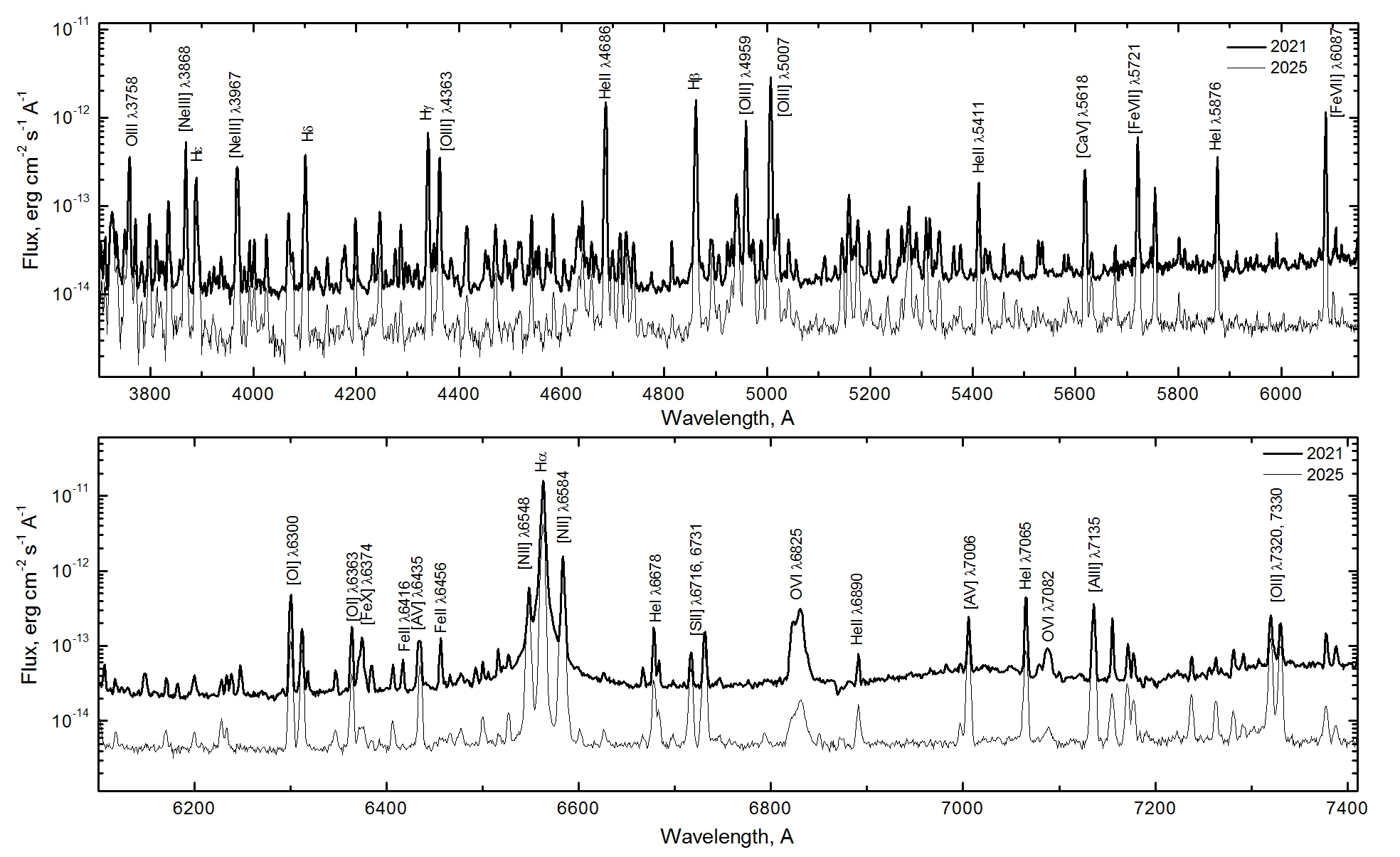}
\caption{Spectra of \hm\ from June 6, 2021 (thick line) and November 18, 2025 (thin line). }
\label{spectra}
   \end{figure*}

Figure~\ref{spectra} compares the spectrum of \hm\ at two epochs: June 6, 2021, and November 18, 2025. Characteristic emission lines are labeled.

% --  --  --  --  --  --  --  -- -Fig. 8 --  --  --  --  --  --  -- --

\subsubsection{Raman-scattered \ion{O}{VI} lines}

The most significant changes during the 2018--2021 event affected the broad emission features in the optical spectral range  --  the $\lambda$6825 and $\lambda$7082 lines, which are produced by Raman scattering of the \ion{O}{VI} resonance doublet ($\lambda\lambda$1032, 1038) by neutral hydrogen (\ion{H}{I}).

In the 1987 spectrum \citep{schmid1990}, as in previous observations, these lines were not detected. The $\lambda$6825 line was first measured in \cite{arkhipova2004} from spectra obtained during 1994--2002. That work also notes that this line was present in an echelle spectrum from 1992. \cite{lee2007} presented and analyzed the profile of the \ion{O}{VI}~$\lambda$6825 line obtained in 2005. The authors modeled the $\lambda$6825 line profile using a Monte Carlo method, assuming that the \ion{O}{VI} emission region is a Keplerian accretion disk around the hot component, while scattering occurs in the slow spherical wind from the Mira variable. Analysis of 2014 observations and modeling of both Raman-scattered \ion{O}{VI} lines 
at $\lambda$6825 and $\lambda$7082 are presented in \cite{lee2016}. That study showed that the flux ratio $F(\lambda6825)/F(\lambda7082)$ of the Raman-scattered lines is a sensitive indicator of the neutral hydrogen column density $N_{\mathrm{H\,I}}$ in the scattering region. Observations of \hm\ in 2014 yielded a flux ratio corresponding to $N_{\mathrm{H\,I}} \sim 10^{22}$~cm$^{-2}$, which is typical of widely separated D-type symbiotic stars.

\begin{figure}
   \centering
   \includegraphics[width=\hsize]{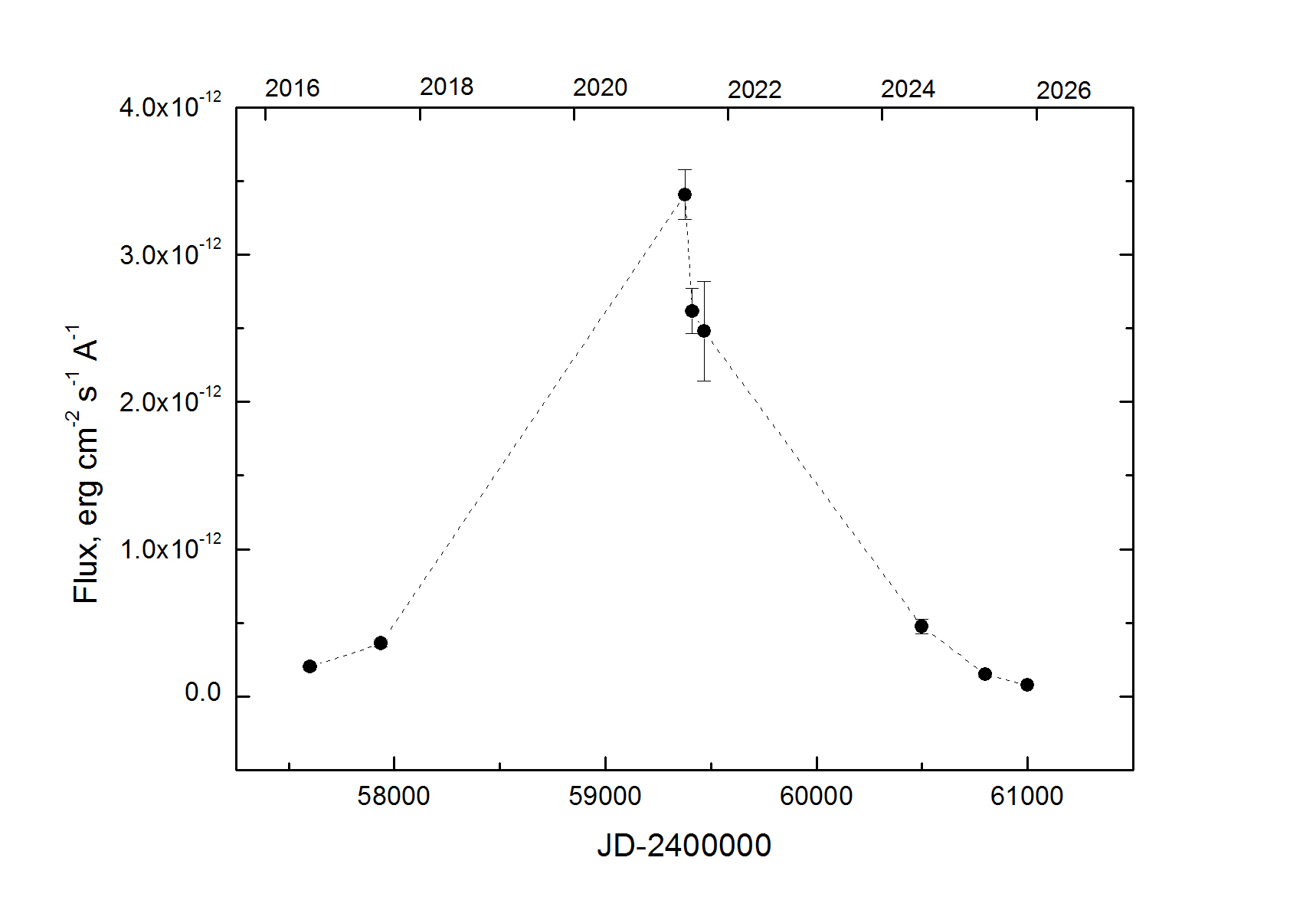}
\caption{Temporal evolution of the absolute flux of the Raman-scattered \ion{O}{VI}~$\lambda$6825 line from 2016 to 2025.}
\label{OVI}
   \end{figure}

Figure~\ref{OVI} shows the evolution of the absolute flux of the Raman-scattered \ion{O}{VI}~$\lambda$6825 line from 2016 to 2025 based on our new spectroscopic observations. Compared to the September 2002 measurement from \cite{arkhipova2004} ($F(\lambda6825) = (1.9 \pm 0.6) \times 10^{-13}$~erg~cm$^{-2}$~s$^{-1}$), the line flux remained nearly unchanged by 2016, but then underwent a sharp increase, reaching an intensity 17 times higher than the 2016 level by 2021. After peaking concurrently with the optical brightness maximum, the flux began to decline: by 2025, it had dropped by a factor of 2.6 relative to the 2016 level. The \ion{O}{VI}~$\lambda$7082 line emerged as a transient feature: its intensity peaked in 2021, and by 2025 the line had virtually disappeared.

% --  --  --  --  --  --  --  -- -Fig. 9 --  --  --  --  --  --  -- --

% --  --  --  --  --  --  --  --  --  --  --  --  --  --  --  -- - 

The behavior of the Raman-scattered lines at $\lambda$6825 and $\lambda$7082 serves as a direct indicator of changes in two system components: the source generating the far ultraviolet (UV) \ion{O}{VI} resonance lines ($\lambda\lambda$1032,1038)  --  that is, the hot component or accretion disk  --  and the extended region of neutral hydrogen surrounding the Mira variable, where Raman scattering occurs. The 17-fold increase in flux of the $\lambda$6825 line during 2018--2021 indicates a qualitative transition, likely caused by a sharp enhancement of accretion onto the hot component, leading to its intense heating. The coincidence of the $\lambda$6825 flux maximum with the optical brightness peak suggests that the source of optical and UV emission (including the \ion{O}{VI} resonance lines) is the same  --  the hot component of the system (WD and/or accretion disk), whose luminosity increased in both wavelength ranges. Finally, the rapid decline of the $\lambda$6825 flux after the peak  --  by a factor of 2.6 by 2025 relative to the 2016 level  --  and the disappearance of the $\lambda$7082 line indicate rapid fading of the ionizing source. Changes in the \ion{H}{I} column density $N_{\mathrm{H\,I}}$ in the giant's slow wind are a much longer-term process, making them unlikely to explain the rapid fading of the Raman lines. Interferometric observations from 2005--2006 \citep{Sacuto2007} showed that the dusty (and therefore gaseous) envelope of the Mira variable is compact and stable on decadal timescales. The mass-loss rate of the Mira component is estimated as $\dot{M} \sim 7 \times 10^{-6}~M_{\odot}$~yr$^{-1}$. To substantially ($\gtrsim 50\%$) change the mean column density in such a compact envelope with characteristic size $R \sim 15$~au and wind expansion velocity $v_{\text{exp}} \sim 25$~km~s$^{-1}$, a timescale of order $\tau \sim R / v_{\text{exp}} \gtrsim 10$--$15$~yr would be required. Such a slow process cannot explain the observed rapid fading of the Raman lines over 4 years (from 2021 to 2025), confirming the dominant role of changes in the hot component rather than in the wind of the cool giant.

\subsubsection{[Fe~X] line}\label{FeX}

Targeted searches for the forbidden [\ion{Fe}{X}]~$\lambda$6374 line in the optical spectrum of \hm\ were conducted during 1983--1985 \citep{wallerstein1988}, using spectra with a wide dynamic range and modern detectors. The [\ion{Fe}{X}] line was not detected in any spectrum. \cite{wallerstein1988} discussed the paradox of this non-detection despite the presence of soft X-ray emission. They suggested that the ionizing radiation spectrum cuts off sharply between 100 and 300~eV, leaving insufficient photons to ionize iron to Fe$^{9+}$.

\begin{figure}
   \centering
   \includegraphics[width=\hsize]{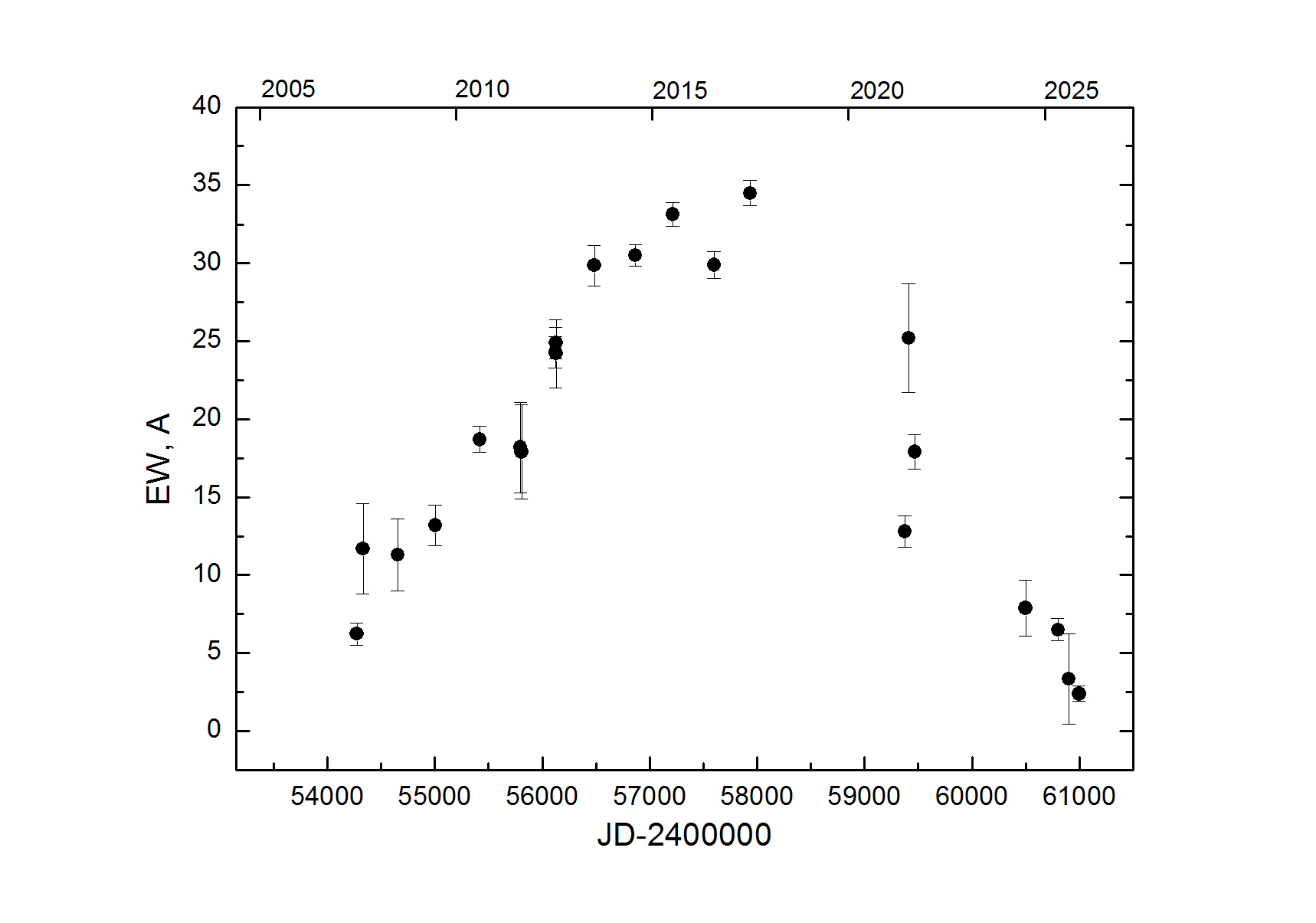}
\caption{Temporal evolution of the equivalent width of [\ion{Fe}{X}]~$\lambda$6374 in 2007--2025. }
\label{EWFeX}
   \end{figure}

Subsequent studies, including our previous work \citep{arkhipova2004}, reported no [\ion{Fe}{X}] detection in the spectrum of \hm. However, our 2016--2025 spectra unambiguously reveal the forbidden [\ion{Fe}{X}]~$\lambda$6374 emission line (Tables~\ref{tab:ZTE} and \ref{tab:TDS}). To determine when the line first appeared, we analyzed the \hm\ spectral archive from the 1.25-m telescope (2003--2015). The earliest signs of [\ion{Fe}{X}]~$\lambda$6374 date to 2007. Due to limited data quality and challenges in absolute flux calibration, we could measure only the equivalent width (EW) of the $\lambda$6374 line for the 2007--2015 spectra. Figure~\ref{EWFeX} and Table~\ref{tab:FeX_EW} show the EW evolution from 2007 to 2025.

\begin{table}[ht]
\centering
\caption{Equivalent width of the [\ion{Fe}{X}]~$\lambda$6374 line, 2007--2025. %DONE Эквивалентная ширина линии [Fe~X] $\lambda 6374$ по наблюдениям 2007-2025~гг.
}
\label{tab:FeX_EW}
\begin{tabular}{cccc}
\hline
Date& JD-2400000 & EW (\AA) & $\sigma_{EW}$ (\AA) \\
\hline
2007-06-22 & 54274.55 & 6.2 & 0.7 \\
2007-08-19 & 54332.47 & 11.7 & 2.9 \\
2008-07-07 & 54655.50 & 11.3 & 2.3 \\
2009-06-23 & 55006.53 & 13.2 & 1.3 \\
2010-08-09 & 55418.42 & 18.7 & 0.8 \\
2011-08-25 & 55799.42 & 18.2 & 2.9 \\
2011-09-01 & 55806.42 & 17.9 & 3.0 \\
2012-07-17 & 56126.44 & 24.3 & 1.0 \\
2012-07-18 & 56127.56 & 24.9 & 1.0 \\
2012-07-19 & 56128.50 & 24.2 & 2.2 \\
2013-07-10 & 56484.49 & 29.9 & 1.3 \\
2014-07-29 & 56868.36 & 30.5 & 0.7 \\
2015-07-12 & 57216.37 & 33.1 & 0.8 \\
2016-07-30 & 57600.46 & 29.9 & 0.9 \\
2017-07-01 & 57936.45 & 34.5 & 0.8 \\
2021-06-10 & 59376.45 & 12.8 & 1.0 \\
2021-07-15 & 59411.34 & 25.2 & 3.5 \\
2021-09-09 & 59467.33 & 17.9 & 1.1 \\
2024-07-07 & 60499.41 & 7.9 & 1.8 \\
2025-05-04 & 60800.47 & 6.5 & 0.7 \\
2025-08-14 & 60902.38 & 3.3 & 2.9 \\
2025-11-18 & 60998.18 & 2.4 & 0.5 \\
\hline
\end{tabular}
\end{table}

% --  --  --  --  --  --  --  -- -Fig. 10 --  --  --  --  --  --  -- --

% --  --  --  --  --  --  --  --  --  --  --  --  --  --  --  -- - 

From 2007 to 2017, the EW increased monotonically from $\sim6$ to $\sim34$~\AA; by 2021 it had decreased to $\sim13$~\AA, and by 2025 it had dropped to minimum values ($\sim2$--$3$~\AA).

Our observations thus trace the [\ion{Fe}{X}] line emergence in 2007, its decade-long growth, and subsequent sharp fading by 2021 -- coincident with the system's optical brightening episode of 2018--2021.

\subsection{[\ion{O}{III}] line formation region: gas diagnostics}

\begin{figure*}
   \centering
   \includegraphics[width=\hsize]{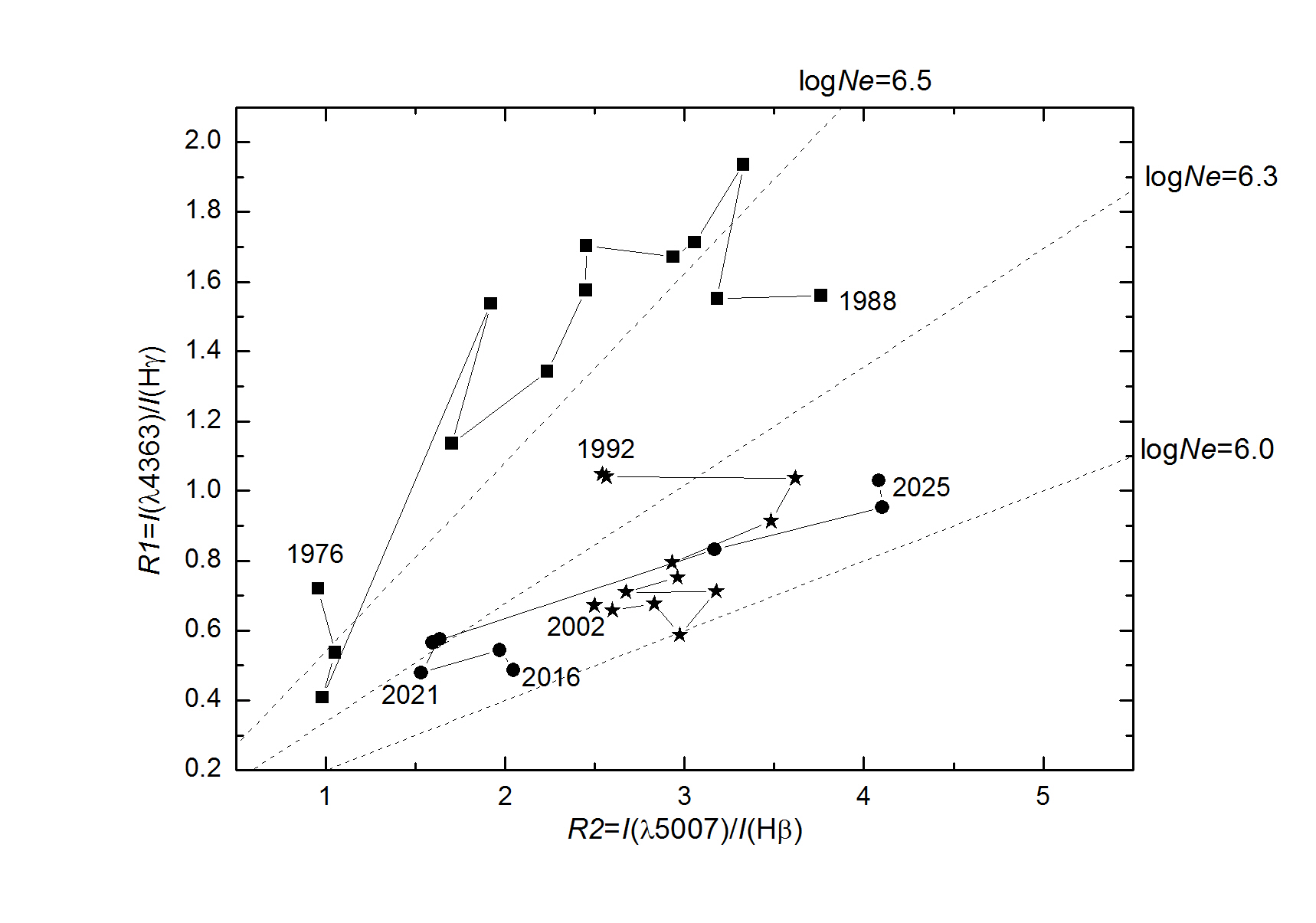}
\caption{$R1$--$R2$ diagram for \hm. Squares correspond to results from \cite{ciatti1977}, \cite{davidson1978}, \cite{blair1981}, \cite{stauffer1984}, \cite{pacheco1989}, \cite{schmid1990}, and \cite{pacheco1992}; stars show measurements from 1992--2002 from \cite{arkhipova2004}; circles represent our new observations from 2016--2025. Numbers on the plot indicate years. Dashed lines show contours of constant electron density $\log N_e$ in the range $6.0 \leq \log N_e \leq 6.5$ from \cite{gutierrez1995}. }

\label{R1-R2}
   \end{figure*}

\cite{gutierrez1995} proposed diagnostic diagrams to distinguish planetary nebulae from symbiotic stars using intensity ratios of the forbidden [\ion{O}{III}] lines ($\lambda\lambda$4363, 5007) and hydrogen Balmer lines. The diagram, constructed with coordinates $R1 = I(\lambda4363)/I(\mathrm{H}\gamma)$ and $R2 = I(\lambda5007)/I(\mathrm{H}\beta)$, effectively separates planetary nebulae from symbiotic stars and distinguishes between S- and D-type symbiotics. In this diagram \hm\ lies in the D-type region. In the same work, based on observations from 1976--1988 compiled from the literature (Table~1 in \cite{gutierrez1995}), the authors analyzed the temporal evolution of \hm's position in the $R1$--$R2$ plane. We extend this analysis with data from \cite{arkhipova2004} and our 2016--2025 measurements. Figure~\ref{R1-R2} shows the object's evolution in the $R1$--$R2$ plane from 1976 to 2025. All values were corrected for interstellar extinction with $E(B-V) = 0.6$.  Although not strictly a diagnostic diagram, in the region where \hm\ evolves ($6.0 \leq \log N_e \leq 6.5$), $R1$ is nearly independent of $N_e$, so its variation reflects changes in $T_e$ alone.

% --  --  --  --  --  --  --  -- -Fig. 11 --  --  --  --  --  --  --  -- 

% --  --  --  --  --  --  --  --  --  --  --  --  --  --  --  -- - 

Several sequential phases can be identified in the evolution of the object on the diagram over the entire observational period. First, during 1976--1988, \hm\ shifts sharply upward and to the right, corresponding to the classical outburst phase of a symbiotic nova. Physically, this is associated with a thermonuclear explosion on the WD surface, which caused a sharp increase in the temperature of the ionizing radiation source and, consequently, a moderate increase in the electron density and ionization degree of the gaseous nebula as well as its heating. Until 1986, \hm\ moves approximately along the line of constant electron density $\log N_e \approx 6.5$.

Then a downward and leftward shift begins along a broken trajectory toward a lower electron density and temperature. This phase corresponds to nebular cooling and post-outburst relaxation: as the ionizing flux declines and the nebula expands, both electron temperature and density drop significantly. During 2016--2021, the electron density increased moderately while the temperature remained stable (decreasing $R2$ at constant $R1$). Starting in 2018, a new activity phase drove \hm\ rapidly upward and rightward on the diagram -- mirroring the post-1975 outburst motion, but at lower $N_e$ values and with a more modest $R1$ increase (by a factor of $\sim2$ compared to the fivefold increase earlier). Since $R1$ variations in this diagram region primarily track $T_e$ changes, the smaller $R1$ increase during 2018--2021 indicates that the electron temperature in the [\ion{O}{III}] line formation region rose less than it did after the 1975 outburst.

Although both activity episodes temporarily elevated electron density in the [\ion{O}{III}] line formation region, the long-term trend across our spectroscopic monitoring shows a gradual decline. Where the object's position in the $R1$--$R2$ plane indicated $N_e \gtrsim 10^{6.5}$~cm$^{-3}$ in the 1980s, by 2025 the data align better with $N_e \gtrsim 10^6$~cm$^{-3}$.

\section{Discussion}

Based on analysis of high-ionization lines and photometric evolution, we suggest that the observed brightening of \hm\ during 2018--2021 in the $U$, $B$, and $V$ bands is likely associated with an increase in the accretion rate onto the WD. The increase in brightness in all photometric bands is probably caused by an increase in the density of ionized gas in the gas continuum formation region (free--bound and free--free), enhancing the emission as $I \propto N_e^2$. The fluxes of emission lines that contribute to the corresponding bands -- \ion{He}{II}~$\lambda$4686 in $B$ and [\ion{Ne}{V}], [\ion{O}{II}]~$\lambda\lambda$3727, 3729 in $U$ -- also increased.

The evolution of the system is also manifested in the IR. Stable periodic oscillations in the $JHKLM$ bands with a period of $\sim$532~days arise from pulsations of the cool Mira component. Slower variations, particularly noticeable in the $J$ band, indicate long-term changes in the dusty envelope. The trend established after the 1975 outburst shows that the dust density reached a maximum approximately 13 years later, then gradually declined. The subsequent $J$-band behavior -- rising to JD 2451000, declining to JD 2454500, then rising again -- suggests a non-stationary process, likely related to variable dust condensation or destruction rates under the radiation field of the hot component.

The 2018--2021 high-activity episode produced significant changes in the system. By 2025, the IR brightness in all bands had declined significantly, reaching the level of the early 1990s, an epoch when the dusty envelope was densest. Although this might suggest a new episode of active dust formation, the lack of reddening in $K-L$ and $L-M$, combined with the decline $J-H$ since 2018, argues against it.  Instead, the clear blueward shift in the $(J-K)-J$ color-magnitude diagram (Fig.~~\ref{fig6}) during 2019--2025 indicates altered physical conditions in the system.

We attribute these changes to enhanced UV/soft X-ray emission from the hot component during 2018-2021, which affected the compact, optically thick silicate envelope around the Mira variable \citep{Sacuto2007}. While the increased high-energy flux could not rapidly alter the envelope's global structure (a decades-long process), it likely modified the thermal balance and grain sizes in its outer layers facing the hot source. This produced two effects: (1) heating and destruction of small grains, reducing the total IR flux; and (2) a change in the shape of near-IR continuum, decreasing the $J-H$ and $J-K$ color indices. Thus, the observed pattern reflects not new dust formation but altered conditions in the pre-existing envelope, driven by changes in the ionizing source. This scenario explains the rapid variations against a steady IR background from the Mira's long-lived dusty wind.

Our recent spectroscopic data allow us to draw several conclusions. Nearly all emission lines in \hm\ responded to the high-activity episode, including hydrogen and helium recombination lines and forbidden lines of highly ionized species. The most striking indicator of enhanced activity was the Raman-scattered \ion{O}{VI}~$\lambda$6825 line, whose flux increased by a factor of 17 by 2021 and then rapidly declined by 2025, which we interpret as evidence for enhanced UV output from the hot component. Contrasting variations in low- and medium-ionization lines ([\ion{O}{III}], [\ion{N}{II}], [\ion{S}{II}]) -- in particular, the strengthening of the [\ion{O}{III}]~$\lambda$4363 line while the [\ion{O}{III}]~$\lambda\lambda$4959, 5007 lines remained constant -- indicate that the ionized envelope responded to the brightening episode primarily through an increase in electron density.

The [\ion{Fe}{X}]~$\lambda$6374 line emerged in 2007 and grew through 2017 despite the fact that the optical brightness declined, which indicates that the ionizing spectrum hardened. This may reflect that the accretion disk settled into a more stationary state, where inner hot regions dominated the output. In contrast, the brightening of 2018--2021 coincided with the weakening of [\ion{Fe}{X}] by 2021. We suggest that the higher accretion rate during this episode caused matter to accumulate on the WD surface, which expanded its outer layers and lowered the photospheric temperature. Consequently, the hard-UV flux needed to produce Fe$^{+9}$ declined, leading to weakening of the [\ion{Fe}{X}] line despite the increase in the system's bolometric luminosity. This picture aligns with WD evolution models at high accretion rates \citep{kuuttila2021}, which also explain the growth of [\ion{Fe}{X}] in 2007--2017: the line luminosity peaks in a narrow temperature range that corresponds to stable nuclear burning (Fig.~7 in \cite{kuuttila2021}). The observed decade-long line growth likely traces the system's approach to this regime as the accretion disk restructured and the hardness of the ionizing radiation increased.

Outbursts are common in classical symbiotic stars and symbiotic novae. In classical symbiotic systems, they arise from enhanced accretion from the red giant, reach amplitudes of a few magnitudes, and are termed Z~And-type outbursts. Such events show rapid brightening followed by a gradual return to quiescence.

Similar outbursts have also been observed in symbiotic novae: AG~Peg \citep{skopal2017}  --  165~yr after its nova-like outburst (amplitude exceeding $1^m.5$); V426~Sge \citep{skopal2020}  --  50~yr after its outburst ($\Delta V\sim1^m.8$). Both systems have well-defined orbital periods --  818~days for AG~Peg and 493~days for V426~Sge -- that persist in their light curves from nova-like to Z~And-type outbursts. The 2018--2021 event in \hm\ differed in several respects: smaller amplitude and no sharp brightening phase.

Moreover, the [\ion{Fe}{X}]~$\lambda$6374 line in \hm\ behaves fundamentally differently from that observed in recurrent symbiotic novae like RS~Oph. The 2010 outburst of V407~Cyg can be considered as an example, where the [\ion{Fe}{X}] line was studied in detail throughout the event \citep{shore2011}. In V407~Cyg, the line appeared within the first week after optical maximum \citep{munari2011}, peaked simultaneously with X-rays ($\sim40$~days), and had disappeared by day~85 after the outburst. This rapid evolution is attributed to shock heating of the gas to temperatures of $\sim1$--$2$~MK as the nova ejecta collide with the dense wind of the Mira variable \citep{shore2011}, with line changes tracing the shock dynamics in an asymmetric circumstellar environment.

In \hm, the picture differs qualitatively. The [\ion{Fe}{X}] line grew over $\sim10$~years, peaked years before optical brightening, and then faded, inconsistent with shock excitation, which requires tight temporal correlation between coronal lines, X-rays, and optical outburst. We interpret the observed behavior as photoionization of the circumstellar environment by hard UV radiation from the accreting WD. The growth of the [\ion{Fe}{X}] line during optical fading reflects spectral hardening as the disk restructured; subsequent weakening is associated with changes in the accretion regime and expansion of the WD atmosphere during the 2018--2021 high-activity episode.

The orbital period of \hm\ significantly exceeds typical values for classical symbiotic stars and some symbiotic novae, which may indicate a different outburst mechanism. The orbital parameters of the system, including the period, have not yet been reliably determined, and the existing estimates are contradictory. Analysis of the proper motion of the radio components N5 and S5 during 1992--1997, interpreted as motion along an elliptical orbit, yielded a separation between the binary components of $\approx25$~au and an orbital period of approximately 90~years \citep{richards1999}. HST imaging provides an estimate of component separation of $\approx50$~au \citep{eyres2001}, which, within standard models, implies an orbital period of several hundred years. However, spectropolarimetry \citep{schmid2000} suggests a period of about 50~years, although the authors emphasize that the quality of the data is insufficient for a definitive conclusion.

If the 1975 nova-like outburst and the 2018-2021 high-activity episode both correspond to periastron passage, their $\sim46$-yr separation could approximate the orbital period. However, this hypothesis conflicts with astrometric constraints and warrants caution. The nature of the 2018–2021 event thus remains ambiguous: it may reflect enhanced periastron accretion, episodic increases in the accretion rate, or weak thermonuclear activity on the WD surface. Resolving this question will require continued multiwavelength monitoring and high-resolution observations.

\section{Conclusion}

Analysis of photometric and spectroscopic observations of \hm\ from 2003 to 2025 yields the following conclusions:

1. After a long-term brightness decline ($\sim0^m.05$~yr$^{-1}$ from 1989 to 2017), \hm\ brightened starting in 2018 and peaked in 2021 ($\Delta V \approx 0^m.3$). By 2025, the $U$, $B$, and $V$ brightness had fallen to the faintest levels in five decades of monitoring. This indicates that after the 2018--2021 high-activity episode, the system has not returned to its pre-brightening state but has continued its evolution with altered parameters. 

2. Near-IR observations confirm Mira-type pulsations with a period of 532~days. During the 2018--2021 optical brightening, IR emission remained largely unchanged except for a modest $J$-band increase. After 2021, all bands ($J$, $H$, $K$, $L$, $M$) faded monotonically, reaching early-1990s levels by 2025 -- when the dusty envelope was densest after the 1975 outburst. The absence of reddening in $K-L$ and $L-M$, combined with bluer $J-H$ and $J-K$ color indices since 2019, rules out new dust formation. Instead, we attribute these changes to the heating and destruction of small dust grains by enhanced UV radiation from the hot component.

3. The 2018--2021 high-activity episode produced significant spectral changes: the fluxes of recombination lines (\ion{H}{I}, \ion{He}{I}, \ion{He}{II}) and high-excitation forbidden lines ([\ion{Ar}{V}], [\ion{Fe}{VI}], [\ion{Fe}{VII}]) increased. The [\ion{Fe}{X}]~$\lambda$6374 line, detected in \hm\ for the first time, began growing in 2007 and peaked in 2017 -- several years before the optical maximum. This timeline indicates photoionization by hard UV radiation from a source (accretion disk or WD) that restructured well before the optical brightening. The [\ion{Fe}{X}] flux declined after 2021, tracking the hot component's fading activity. This behavior matches models, where high accretion rates expand the WD's outer layers and lower its photospheric temperature \citep{kuuttila2021}.

4. The flux of the Raman-scattered \ion{O}{VI}~$\lambda$6825 line increased 17-fold during 2018--2021, directly tracing enhanced ionizing radiation from the hot component. The system's position on the [\ion{O}{III}] diagnostic diagram shifted toward higher electron temperatures, marking the onset of a new heating cycle in the gaseous envelope.

5. We interpret the 2018--2021 event as resulting from enhanced accretion from the red giant onto the WD. The 1975 outburst and 2018--2021 episode may both correspond to periastron passage; if so, the $\sim$46-yr interval would approximate the orbital period. However, this hypothesis requires further verification. 

This paper presents a qualitative analysis of new observational data for \hm\ from 2003 to 2025. Our results reveal substantial changes in the photometric and spectral behavior of the system. However, a more complete understanding of the underlying physical processes will require quantitative modeling. In future work, we plan to determine fundamental system parameters and simulate the spectral energy distribution,  accounting for contributions from various system components.

\hm\ remains a unique laboratory for studying the late stages of symbiotic nova evolution. Clarifying the processes driving recurrent heating and long-period activity will require further observations across all accessible wavelength ranges.

\bigskip
\acknowledgments

The study was conducted under the state assignment of the Lomonosov Moscow State University.

\bigskip
\bibliographystyle{aspb1}
\bibliography{HMSge_arxiv}
\newpage

\widetext  
\section*{Appendix}
\setcounter{table}{0}
\renewcommand\thetable{A\arabic{table}}

\begin{longtable*}{cccc}
\caption{$UBV$ photometry for \hm, 2003--2025.}
\label{tab:UBV}\\
\toprule
JD-2400000 & $U$ & $B$ & $V$ \\
\midrule
\endfirsthead
\toprule
JD-2400000 & $U$ & $B$ & $V$ \\
\midrule
\endhead
\bottomrule
\endfoot
57213 & 12.14 & 13.09 & 12.56 \\
57216 & 12.11 & 13.05 & 12.53 \\
57217 & 12.12 & 13.09 & 12.50 \\
57249 & 12.13 & 13.08 & 12.51 \\
57272 & 12.12 & 13.07 & 12.53 \\
57299 & 12.15 & 13.10 & 12.54 \\
57517 & 12.20 & 13.15 & 12.57 \\
57549 & 12.23 & 13.13 & 12.60 \\
57573 & 12.16 & 13.11 & 12.53 \\
57584 & 12.17 & 13.12 & 12.54 \\
57638 & 12.25 & 13.12 & 12.55 \\
57640 & 12.23 & 13.15 & 12.55 \\
57935 & 12.23 & 13.19 & 12.59 \\
57958 & 12.26 & 13.17 & 12.59 \\
57959 & 12.29 & 13.21 & 12.56 \\
57980 & 12.33 & 13.18 & 12.57 \\
58013 & 12.28 & 13.20 & 12.60 \\
58341 & 12.16 & 13.05 & 12.45 \\
58344 & 12.19 & 13.05 & 12.46 \\
58351 & 12.18 & 13.04 & 12.45 \\
58366 & 12.16 & 13.04 & 12.45 \\
58600 & 12.04 & 12.95 & 12.40 \\
58607 & 12.03 & 12.94 & 12.41 \\
58617 & 12.04 & 12.92 & 12.38 \\
58635 & 12.02 & 12.90 & 12.37 \\
58638 & 12.04 & 12.93 & 12.39 \\
58647 & 12.00 & 12.92 & 12.39 \\
58658 & 12.04 & 12.89 & 12.36 \\
58661 & 11.99 & 12.90 & 12.37 \\
58677 & 12.00 & 12.90 & 12.34 \\
58672 & 12.02 & 12.91 & 12.37 \\
58691 & 12.00 & 12.90 & 12.34 \\
58701 & 11.99 & 12.89 & 12.37 \\
58704 & 12.00 & 12.92 & 12.36 \\
58719 & 12.00 & 12.88 & 12.34 \\
58729 & 12.01 & 12.89 & 12.36 \\
58730 & 12.00 & 12.91 & 12.35 \\
58753 & 12.00 & 12.91 & 12.36 \\
58760 & 12.01 & 12.88 & 12.36 \\
59043 & 12.03 & 12.90 & 12.32 \\
59060 & 12.00 & 12.90 & 12.33 \\
59078 & 11.99 & 12.93 & 12.33 \\
59079 & 12.01 & 12.90 & 12.36 \\
59087 & 12.00 & 12.86 & 12.35 \\
59110 & 11.98 & 12.90 & 12.34 \\
59145 & 11.98 & 12.92 & 12.38 \\
59339 & 11.92 & 12.88 & 12.33 \\
59374 & 11.94 & 12.86 & 12.28 \\
59377 & 11.92 & 12.85 & 12.27 \\
59410 & 11.94 & 12.84 & 12.27 \\
59429 & 11.94 & 12.85 & 12.26 \\
59339 & 11.92 & 12.88 & 12.33 \\
59374 & 11.94 & 12.86 & 12.28 \\
59377 & 11.92 & 12.85 & 12.27 \\
59410 & 11.94 & 12.84 & 12.27 \\
59429 & 11.94 & 12.85 & 12.26 \\
59732 & 12.10 & 13.03 & 12.44 \\
59760 & 12.06 & 13.03 & 12.45 \\
59765 & 12.08 & 13.05 & 12.47 \\
59766 & 12.05 & 13.04 & 12.46 \\
59787 & 12.07 & 13.07 & 12.47 \\
59796 & 12.04 & 13.06 & 12.46 \\
59821 & 12.10 & 13.14 & 12.55 \\
59732 & 12.10 & 13.03 & 12.44 \\
59765 & 12.08 & 13.05 & 12.47 \\
59766 & 12.05 & 13.04 & 12.46 \\
59787 & 12.07 & 13.07 & 12.48 \\
59796 & 12.04 & 13.06 & 12.46 \\
59851 & 12.10 & 13.14 & 12.55 \\
60083 & 12.32 & 13.39 & 12.76 \\
60116 & 12.36 & 13.39 & 12.70 \\
60120 & 12.37 & 13.42 & 12.73 \\
60208 & 12.45 & 13.49 & 12.76 \\
60223 & 12.43 & 13.51 & 12.80 \\
60469 & 12.53 & 13.62 & 12.84 \\
60494 & 12.51 & 13.62 & 12.87 \\
60534 & 12.58 & 13.67 & 12.91 \\
60536 & 12.54 & 13.68 & 12.95 \\
60863 & 12.84 & 14.04 & 13.13 \\
60877 & 12.82 & 13.99 & 13.14 \\
60878 & 12.80 & 13.98 & 13.11 \\

\end{longtable*}

\begin{longtable*}{cccccc}
\caption{$JHKLM$ photometry for \hm, 2009-2025.}
\label{tab:JHKLM}\\
\toprule
JD-2400000 & $J$ & $H$ & $K$ & $L$ & $M$\\
\midrule
\endfirsthead
\toprule
JD-2400000 & $J$ & $H$ & $K$ & $L$ & $M$\\
\midrule
\endhead
\bottomrule
\endfoot

54982.5 & 8.34 & 6.38 & 4.56 & 2.33 & 1.65 \\
54994.5 & 8.32 & 6.31 & 4.49 & 2.26 & 1.65 \\
55014.4 & 8.02 & 6.07 & 4.27 & 2.09 & 1.58 \\
55021.4 & 7.91 & 5.99 & 4.16 & 1.98 & 1.49 \\
55047.4 & 7.59 & 5.69 & 3.92 & 1.78 & 1.20\\
55056.4 & 7.51 & 5.64 & 3.89 & 1.76 & 1.20 \\
55113.2 & 7.20 & 5.44 & 3.72 & 1.66 & 1.20 \\
55145.1 & 7.10 & 5.30 & 3.68 & 1.65 & 1.15 \\
55318.5 & 7.87 & 5.97 & 4.28 & 2.16 & 1.63 \\
55408.4 & 8.50 & 6.48 & 4.65 & 2.44 & 1.91 \\
55431.3 & 8.51 & 6.50 & 4.62 & 2.43 & 1.76 \\
55434.3 & 8.44 & 6.44 & 4.65 & 2.42 & 1.81 \\
55463.2 & 8.44 & 6.38 & 4.57 & 2.38 & 1.73 \\
55493.2 & 8.29 & 6.28 & 4.50 & 2.31 & 1.73 \\
55503.2 & 8.20 & 6.21 & 4.44 & 2.28 & 1.75 \\
55692.5 & 6.95 & 5.21 & 3.64 & 1.64 & 1.06 \\
55729.5 & 6.95 & 5.24 & 3.68 & 1.68 & 1.31 \\
55751.4 & 7.01 & 5.31 & 3.74 & 1.71 & 1.14 \\
55759.5 & 7.08 & 5.34 & 3.75 & 1.73 & 1.26 \\
55780.4 & 7.22 & 5.49 & 3.87 & 1.82 & 1.30 \\
55784.4 & 7.26 & 5.50 & 3.90 & 1.86 & 1.34 \\
55794.4 & 7.34 & 5.55 & 3.96 & 1.89 & 1.37 \\
55811.3 & 7.49 & 5.70 & 4.06 & 1.99 & 1.42 \\
55820.3 & 7.57 & 5.77 & 4.10 & 2.02 & 1.54 \\
55843.2 & 7.84 & 5.96 & 4.26 & 2.14 & 1.53 \\
55870.2 & 8.05 & 6.16 & 4.41 & 2.26 & 1.66 \\
56059.5 & 8.20 & 6.25 & 4.39 & 2.21 & 1.73 \\
56084.5 & 7.91 & 6.04 & 4.27 & 2.11 & 1.60 \\
56092.5 & 7.74 & 5.91 & 4.17 & 2.04 & 1.51 \\
56111.4 & 7.42 & 5.67 & 3.96 & 1.88 & 1.35 \\
56136.4 & 7.20 & 5.50 & 3.83 & 1.78 & 1.13 \\
56197.3 & 6.97 & 5.29 & 3.69 & 1.68 & 1.05 \\
56224.2 & 6.94 & 5.19 & 3.62 & 1.69 & 1.06 \\
56434.5 & 8.23 & 6.28 & 4.52 & 2.37 & 1.81 \\
56470.5 & 8.35 & 6.41 & 4.61 & 2.47 & 1.95 \\
56473.4 & 8.38 & 6.42 & 4.61 & 2.49 & 1.92 \\
56485.5 & 8.42 & 6.42 & 4.64 & 2.48 & 1.98 \\
56497.4 & 8.46 & 6.46 & 4.65 & 2.49 & 1.96 \\
56518.4 & 8.40 & 6.42 & 4.63 & 2.43 & 1.84 \\
56579.2 & 8.32 & 6.34 & 4.53 & 2.38 & 1.78 \\
56824.5 & 7.12 & 5.33 & 3.74 & 1.76 & 1.27 \\
56848.4 & 7.16 & 5.34 & 3.77 & 1.81 & 1.35 \\
56875.4 & 7.18 & 5.41 & 3.84 & 1.88 & 1.32 \\
56883.4 & 7.24 & 5.45 & 3.89 & 1.92 & 1.47 \\
56914.3 & 7.42 & 5.57 & 4.02 & 2.11 & 1.57 \\
56937.2 & 7.63 & 5.77 & 4.18 & 2.16 & 1.55 \\
56967.2 & 7.86 & 5.97 & 4.33 & 2.30 & 1.70 \\
57205.4 & 7.35 & 5.60 & 3.91 & 1.87 & 1.33 \\
57234.4 & 7.20 & 5.48 & 3.85 & 1.81 & 1.33 \\
57261.4 & 7.14 & 5.44 & 3.81 & 1.79 & 1.28 \\
57268.3 & 7.10 & 5.40 & 3.78 & 1.78 & 1.25 \\
57287.3 & 6.98 & 5.28 & 3.68 & 1.70 & 1.25 \\
57525.5 & 7.88 & 6.12 & 4.48 & 2.46 & 1.93 \\
57561.5 & 7.91 & 6.14 & 4.54 & 2.47 & 2.03 \\
57584.4 & 7.90 & 6.14 & 4.52 & 2.46 & 1.90 \\
57592.4 & 7.88 & 6.12 & 4.51 & 2.46 & 2.02 \\
57618.4 & 7.85 & 6.06 & 4.45 & 2.40 & 1.94 \\
57637.4 & 7.81 & 6.05 & 4.44 & 2.44 & 1.94 \\
57646.3 & 7.76 & 6.03 & 4.42 & 2.39 & 1.83 \\
57916.5 & 6.91 & 5.33 & 3.90 & 1.94 & 1.44 \\
57943.5 & 6.98 & 5.43 & 4.04 & 2.06 & 1.47 \\
57968.5 & 7.10 & 5.55 & 4.14 & 2.23 & 1.73 \\
57973.4 & 7.06 & 5.51 & 4.14 & 2.22 & 1.76 \\
58000.4 & 7.20 & 5.66 & 4.22 & 2.34 & 1.78 \\
58242.6 & 7.25 & 5.72 & 4.27 & 2.33 & 1.89 \\
58303.5 & 6.81 & 5.39 & 4.04 & 2.11 & 1.71 \\
58330.4 & 6.75 & 5.37 & 4.00 & 2.05 & 1.69 \\
58352.4 & 6.73 & 5.36 & 3.97 & 2.07 & 1.49 \\
58363.4 & 6.72 & 5.35 & 3.96 & 2.04 & 1.60 \\
58386.2 & 6.71 & 5.32 & 3.95 & 2.02 & 1.56 \\
58415.2 & 6.78 & 5.33 & 4.00 & 2.08 & 1.54 \\
58448.1 & 6.76 & 5.36 & 4.01 & 2.10 & 1.6 \\
58631.5 & 7.66 & 6.05 & 4.58 & 2.66 & -- \\
58653.5 & 7.70 & 6.08 & 4.61 & 2.66 & 2.02 \\
58678.4 & 7.71 & 6.10 & 4.61 & 2.67 & 2.18 \\
58709.3 & 7.72 & 6.08 & 4.55 & 2.59 & 1.98 \\
58721.3 & 7.68 & 6.09 & 4.55 & 2.58 & 2.06 \\
58736.3 & 7.73 & 6.11 & 4.58 & 2.60 & 2.00 \\
58766.3 & 7.69 & 6.07 & 4.54 & 2.56 & 2.05 \\
58797.2 & 7.38 & 5.86 & 4.35 & 2.35 & 1.78 \\
58972.5 & 6.96 & 5.58 & 4.17 & 2.17 & 1.60 \\
59026.5 & 7.13 & 5.73 & 4.30 & 2.32 & 1.76 \\
59041.4 & 7.24 & 5.79 & 4.41 & 2.38 & 1.84 \\
59060.4 & 7.31 & 5.90 & 4.46 & 2.44 & 2.02 \\
59076.4 & 7.37 & 5.93 & 4.52 & 2.44 & 1.79 \\
59095.3 & 7.51 & 6.03 & 4.57 & 2.56 & 1.99 \\
59124.2 & 7.72 & 6.20 & 4.70 & 2.68 & 2.14 \\
59166.2 & 7.89 & 6.31 & 4.78 & 2.77 & 2.23 \\
59376.5 & 6.94 & 5.55 & 4.10 & 2.15 & 1.66 \\
59386.4 & 6.89 & 5.52 & 4.08 & 2.13 & 1.70 \\
59415.5 & 6.83 & 5.43 & 3.98 & 2.04 & 1.56 \\
59444.5 & 6.79 & 5.39 & 4.00 & 2.02 & 1.53 \\
59452.3 & 6.82 & 5.43 & 3.99 & 2.06 & 1.58 \\
59517.2 & 6.94 & 5.52 & 4.08 & 2.12 & 1.60 \\
59768.4 & 8.22 & 6.48 & 4.86 & 2.79 & 2.22 \\
59780.4 & 8.20 & 6.47 & 4.83 & 2.79 & 2.24 \\
59867.2 & 7.72 & 6.15 & 4.61 & 2.61 & 2.08 \\
59893.2 & 7.39 & 5.89 & 4.39 & 2.38 & 1.87 \\
60123.5 & 7.50 & 5.84 & 4.33 & 2.34 & 1.84 \\
60154.4 & 7.74 & 6.08 & 4.52 & 2.14 & 2.02 \\
60184.4 & 7.94 & 6.30 & 4.71 & 2.66 & 2.26 \\
60211.2 & 8.16 & 6.45 & 4.84 & 2.74 & 2.23 \\
60247.2 & 8.39 & 6.64 & 4.99 & 2.91 & 2.30 \\
60447.5 & 7.40 & 5.87 & 4.32 & 2.28 & 1.82 \\
60462.4 & 7.24 & 5.74 & 4.21 & 2.20 & 1.80 \\
60485.5 & 7.08 & 5.61 & 4.10 & 2.12 & 1.66 \\
60512.4 & 7.07 & 5.57 & 4.06 & 2.12 & 1.71 \\
60538.4 & 7.04 & 5.52 & 4.05 & 2.04 & 1.61 \\
60567.3 & 7.04 & 5.51 & 4.05 & 2.08 & 1.66 \\
60622.2 & 7.13 & 5.61 & 4.15 & 2.17 & 1.81 \\
60818.5 & 8.68 & 6.96 & 5.31 & 3.08 & 2.57 \\
60833.5 & 8.75 & 7.00 & 5.34 & 3.06 & 2.56 \\

\end{longtable*}

\newpage

\begin{longtable*}{ccccccc}
\caption{Absolute emission line fluxes in the spectrum of \hm\ from observations obtained with the 1.25-m telescope. Fluxes are given in units of $10^{-13}$~erg~cm$^{-2}$~s$^{-1}$. 
} \label{tab:ZTE} \\
\toprule
$\lambda_{\text{lab}}$ (\AA) & Ion & $F$(2016-07-30) & $F$(2017-07-01) & $F$(2021-07-15) & $F$(2021-09-09) & $F$(2024-07-07) \\
\midrule
\endfirsthead

\caption*{Continuation of Table~\ref{tab:ZTE}. Absolute fluxes are given in units of $10^{-13}$~erg~cm$^{-2}$~s$^{-1}$. 
} \\
\toprule
$\lambda_{\text{lab}}$ (\AA) & Ion & $F$(2016-07-30) & $F$(2017-07-01) & $F$(2021-07-15) & $F$(2021-09-09) & $F$(2024-07-07) \\
\midrule
\endhead

\bottomrule
\endfoot

4101.76 & H I & 8.01 & 7.60 & 12.06 & 12.23 & -- \\
4340.47 & H I & 16.67 & 15.46 & 23.98 & 20.30 & 12.46 \\
4363.21 & [O III] & 8.21 & 8.51 & 13.99 & 11.63 & 10.50 \\
4416.5bl & [Fe II] & -- & -- & 2.36 & 1.92 & -- \\
4471.48 & He I & 0.63 & 0.51 & 2.00 & 1.93 & -- \\
4541.59 & He II & 0.96 & 1.03 & 2.53 & 2.38 & -- \\
4582.83 & Fe II & -- & 1.22 & -- & -- & -- \\
4685.68 & He II & 35.66 & 35.17 & 60.86 & 51.12 & 27.78 \\
4861.33 & H I & 50.29 & 49.98 & 68.31 & 56.82 & 34.05 \\
4940.68 & -- & 7.88 & 8.12 & 10.58 & 8.41 & 8.40 \\
4958.92 & [O III] & 41.08 & 40.38 & 45.43 & 38.60 & 43.67 \\
5006.84 & [O III] & 125.90 & 120.50 & 136.70 & 110.90 & 131.90 \\
5145.75 & [Fe VI] & 0.82 & 0.67 & 1.21 & 1.31 & 1.49 \\
5158.01 & [Fe II] & 1.65 & 2.07 & 5.98 & 5.25 & 3.11 \\
5176.04 & [Fe VI] & 1.60 & 1.86 & 3.51 & 3.51 & 2.97 \\
5197.90 & [N I] & 0.38 & 0.53 & 1.58 & 1.47 & -- \\
5234.28 & [Fe VI] & 0.22 & -- & 1.57 & -- & 0.35 \\
5275.82 & [Ni II] & 1.97 & 2.25 & 5.76 & 4.73 & 3.38 \\
5309.02 & [Ca V] & 1.38 & 1.95 & 5.45 & 4.90 & 2.61 \\
5335.18 & [Fe VI] & -- & 0.54 & 2.50 & 2.24 & 0.64 \\
5412.00 & He II / [Fe III] & 4.45 & 4.70 & 7.93 & 6.78 & 3.39 \\
5427.82 & [Fe II] & -- & -- & -- & -- & 0.63 \\
5460.50 & Fe II & -- & -- & -- & 0.75 & 0.46 \\
5619.00 & [Ca VII] & 6.94 & 6.85 & 11.18 & 10.30 & 7.54 \\
5631.07 & [Fe VI] & -- & -- & -- & -- & 0.62 \\
5676.95 & [Fe VI] & -- & -- & -- & -- & 0.57 \\
5721.15 & [Fe VII] & 10.39 & 10.90 & 17.00 & 16.20 & 18.00 \\
5754.64 & [N II] & 4.39 & 4.42 & 4.10 & 3.65 & 3.01 \\
5875.67 & He I & 4.53 & 4.65 & 8.77 & 8.09 & 4.17 \\
6086.92 & [Fe VII] & 18.92 & 19.75 & 29.02 & 27.05 & 30.82 \\
6300.30 & [O I] & 7.15 & 8.13 & 11.16 & 10.28 & 4.70 \\
6312.06 & [S III] & 3.75 & 4.26 & 3.75 & 3.68 & 3.04 \\
6363.78 & [O I] & 2.45 & 2.41 & 3.80 & 3.61 & 1.56 \\
6374.53 & [Fe X] & 4.12 & 5.94 & 4.05 & 3.52 & 0.67 \\
6435.10 & [Ar V] & 2.01 & 2.46 & 3.50 & 3.31 & 1.98 \\
6548.03 & [N II] & -- & -- & 31.44 & 24.70 & 19.32 \\
6562.82 & H I & 329.90 & 404.70 & 511.00 & 473.70 & 209.10 \\
6583.45 & [N II] & 45.72 & 59.32 & 56.11 & 53.78 & 51.34 \\
6678.15 & He I & 1.68 & 1.89 & 4.01 & 3.68 & 1.62 \\
6716.47 & [S II] & 1.45 & 1.58 & 1.63 & 1.73 & 2.11 \\
6730.85 & [S II] & 2.71 & 3.24 & 3.46 & 3.41 & 4.13 \\
6825.00 & O VI & 2.01 & 3.59 & 26.16 & 24.80 & 4.74 \\
6891.00 & He II & 0.87 & 0.87 & 1.19 & 1.18 & 0.64 \\
7005.83 & [Ar V] & 5.29 & 5.39 & 5.54 & 5.06 & 4.07 \\
7065.28 & He I & 4.19 & 4.28 & 9.60 & 8.70 & 3.56 \\
7135.78 & [Ar III] & 9.32 & 8.69 & 8.47 & 8.07 & 8.16 \\
7155.17 & [Fe II] & 0.64 & 0.96 & 4.41 & 3.93 & 1.25 \\
7171.62 & [Ar IV] & 0.84 & 1.77 & -- & 3.26 & 1.07 \\
7237.17 & C II & 0.69 & 0.65 & -- & 0.72 & 0.70 \\
7263.33 & [Ar IV] & 0.61 & 0.59 & -- & 0.53 & 0.57 \\
7281.35 & He I & 0.50 & -- & -- & 0.90 & -- \\
7291.00 & -- & -- & -- & -- & 1.11 & -- \\
7319.40 & [O II] & 7.47 & 8.10 & 7.34 & 6.27 & 5.41 \\
7330.70 & [O II] & 6.22 & 5.50 & 5.52 & 4.62 & 4.68 \\
7377.83 & [Ni II] & -- & 0.53 & 2.36 & 1.93 & 0.85 \\
7388.40 & [Fe II] & -- & -- & 1.22 & 1.04 & 0.38 \\
7411.61 & [Ni II] & -- & -- & 0.83 & 0.75 & 0.27 \\
7452.56 & [Fe II] & -- & 0.29 & 1.29 & 1.34 & 0.35 \\
7530.83 & [Cl IV] & 0.72 & 0.27 & -- & -- & 0.22 \\
7593.00 & -- & 2.65 & 3.17 & 1.95 & 2.44 & 2.37 \\
7712.28 & Fe II & 2.13 & 2.60 & 3.31 & 3.81 & -- \\
7751.43 & [Ar III] & 2.36 & 2.55 & 1.72 & 1.72 & 2.10 \\
7866.56 & Fe II & -- & 0.39 & 1.71 & -- & -- \\
7967.00 & -- & 2.10 & 2.29 & -- & -- & -- \\
8046.10 & [Cl IV] & 0.62 & 0.65 & -- & -- & 0.62 \\
8237.92 & Fe II & 3.00 & 2.52 & -- & -- & -- \\
8445.00 & O I & 5.02 & 9.53 & -- & -- & -- \\
8467.26 & H I & 0.72 & -- & -- & -- & -- \\
8502.49 & H I & 0.79 & -- & -- & -- & -- \\
8545.38 & H I & 0.96 & -- & -- & -- & -- \\
8598.39 & H I & 1.23 & -- & -- & -- & -- \\
8665.02 & H I & 1.53 & -- & -- & -- & -- \\
8750.48 & H I & 1.74 & -- & -- & -- & -- \\
8862.79 & H I & 2.31 & -- & -- & -- & -- \\
8926.66 & Fe II & 1.54 & -- & -- & -- & -- \\
9014.91 & H I & 4.36 & -- & -- & -- & -- \\
9068.60 & [S III] & -- & 15.13 & -- & -- & 15.09 \\
9530.60 & [S III] & -- & 41.99 & -- & -- & -- \\
\end{longtable*}

\begin{longtable*}{ccccc}
\caption{Absolute emission line fluxes in the spectrum of \hm\ from observations obtained with the TDS spectrograph. Fluxes are given in units of $10^{-13}$~erg~cm$^{-2}$~s$^{-1}$. 
}\label{tab:TDS}\\
\toprule
$\lambda_{\text{lab}}$ (\AA) & Ion & $F$ (2021-06-10) & $F$ (2025-05-04) & $F$ (2025-11-18) \\
\midrule
\endfirsthead
\caption*{Continuation of Table~\ref{tab:TDS}. Absolute fluxes are given in units of $10^{-13}$~erg~cm$^{-2}$~s$^{-1}$. 
} \\
\toprule
$\lambda_{\text{lab}}$ (\AA) & Ion & $F$ (2021-06-10) & $F$ (2025-05-04) & $F$ (2025-11-18) \\
\midrule
\endhead
\bottomrule
\endfoot
3585.99 & -- & 7.77 & 8.16 & 8.28 \\
3727.50 & [O II]bl & 6.35 & 5.65 & 4.78 \\
3757.40 & [Fe VI] & 12.39 & 11.18 & 11.82 \\
3835.40 & H I & 4.38 & -- & -- \\
3868.76 & [Ne III] & 18.60 & 16.67 & 17.35 \\
3889.02 & H I+He I & 8.13 & 3.65 & 3.65 \\
3968.72 & [Ne III]+H I & 14.60 & 8.39 & 8.64 \\
4068.60 & [S II] & 3.84 & 2.15 & 2.16 \\
4076.35 & [S II] & 1.10 & 0.72 & 0.76 \\
4101.76 & H I & 14.58 & 5.47 & 5.29 \\
4199.10 & He II & 2.59 & 0.94 & 0.72 \\
4246.23 & Fe II & 4.11 & 1.16 & 0.89 \\
4340.47 & H I & 25.20 & 9.94 & 9.54 \\
4363.21 & [O III] & 12.21 & 9.60 & 9.95 \\
4416.50 & [Fe II]bl & 2.84 & 0.29 & 0.26 \\
4471.48 & He I & 2.01 & 0.67 & 0.68 \\
4541.59 & He II & 2.36 & 0.70 & 0.65 \\
4582.83 & Fe II & 2.84 & 0.31 & 0.16 \\
4634.14 & N III & 1.71 & 0.53 & 0.50 \\
4640.64 & N III & 3.42 & 1.15 & 0.99 \\
4656.98 & Fe II & 1.05 & 0.57 & 0.59 \\
4685.68 & He II & 60.32 & 23.20 & 23.16 \\
4698.82 & [Fe VII] & 0.91 & 0.88 & 0.99 \\
4713.14 & He I & 1.71 & 1.17 & 1.14 \\
4725.62 & [Ne IV] & 2.31 & 1.30 & 1.25 \\
4740.12 & [Ar IV] & 1.08 & 1.20 & 1.26 \\
4774.73 & [Fe II] & 0.20 & -- & -- \\
4814.54 & [Fe II] & 1.15 & -- & -- \\
4861.33 & H I & 68.43 & 29.50 & 29.61 \\
4891.57 & -- & 4.61 & 1.51 & 1.63 \\
4940.68 & [Ca VII] & 9.48 & 5.34 & 5.10 \\
4958.92 & [O III] & 40.17 & 48.50 & 48.24 \\
4988.40 & [Fe VII] & 1.40 & 1.71 & 1.88 \\
5006.84 & [O III] & 128.10 & 148.00 & 147.90 \\
5145.75 & [Fe VI] & 1.18 & 1.35 & 1.55 \\
5158.01 & [Fe II] & 5.37 & 2.79 & 2.94 \\
5176.04 & [Fe VI] & 2.85 & 2.77 & 2.94 \\
5197.90 & [N I] & 1.66 & 0.26 & 0.30 \\
5275.82 & [Ni II] & 3.85 & 2.76 & 2.86 \\
5309.11 & [Ca V] & 2.66 & 2.27 & 2.41 \\
5316.77 & Fe II & 2.65 & -- & -- \\
5411.48 & He II & 7.60 & 2.41 & 2.15 \\
5425.27 & Fe II & 0.73 & -- & -- \\
5527.35 & [Fe II] & 0.98 & -- & -- \\
5534.89 & Fe II & 0.99 & -- & -- \\
5577.34 & [O I] & 0.49 & 0.07 & 0.08 \\
5618.75 & [Ca VII] & 11.10 & 5.43 & 4.78 \\
5631.07 & [Fe VI] & 0.62 & -- & -- \\
5676.95 & [Fe VI] & 0.48 & 0.59 & 0.72 \\
5721.15 & [Fe VII] & 19.16 & 16.20 & 19.11 \\
5754.64 & [N II] & 4.38 & 2.85 & 2.94 \\
5875.67 & He I & 10.10 & 3.31 & 3.33 \\
6086.92 & [Fe VII] & 33.17 & 28.50 & 27.43 \\
6300.30 & [O I] & 13.47 & 3.58 & 3.55 \\
6312.06 & [S III] & 4.21 & 2.69 & 2.83 \\
6363.78 & [O I] & 4.55 & 1.19 & 1.20 \\
6374.53 & [Fe X] & 3.63 & 0.21 & 0.11 \\
6435.10 & [Ar V] & 4.33 & 1.63 & 1.71 \\
6416.93 & Fe II & 0.93 & -- & -- \\
6455.15 & Fe II & 2.61 & -- & -- \\
6548.03 & [N II] & 17.06 & 17.20 & 17.17 \\
6562.82 & H I & 608.70 & 167.00 & 166.80 \\
6583.45 & [N II] & 52.80 & 51.70 & 51.74 \\
6666.87 & [Ni II] & 0.66 & 0.06 & 0.08 \\
6678.15 & He I & 3.95 & 0.95 & 0.95 \\
6683.21 & He II & 0.99 & 0.27 & 0.21 \\
6716.47 & [S II] & 1.69 & 1.66 & 1.68 \\
6730.85 & [S II] & 4.09 & 3.39 & 3.52 \\
6825.00 & O VI & 34.07 & 1.51 & 0.77 \\
6890.80 & He II & 1.08 & 0.36 & 0.37 \\
7005.83 & [Ar V] & 6.08 & 3.72 & 4.10 \\
7065.28 & He I & 11.84 & 2.63 & 2.55 \\
7087.85 & O VI & 5.21 & -- & -- \\
7135.78 & [Ar III] & 9.90 & 7.22 & 7.63 \\
7155.17 & [Fe II] & 5.36 & 0.63 & 0.63 \\
7237.17 & C II & 0.90 & 0.56 & 0.58 \\
7291.00 & -- & 0.78 & -- & -- \\
7319.40 & [O II] & 6.76 & 4.07 & 4.18 \\
7330.83 & [O II] & 5.34 & 3.31 & 3.73 \\
7377.83 & [Ni II] & 2.79 & 0.37 & 0.39 \\
7388.40 & [Fe II] & 1.03 & 0.10 & 0.49 \\
7411.61 & [Ni II] & 1.10 & -- & -- \\
7452.56 & [Fe II] & 1.31 & -- & 0.26 \\

\end{longtable*}

\end{document}